\documentclass[aps,prl,preprintnumbers,amsmath,amssymb,latexsym,nofootinbib,array,enumerate,letter,twocolumn]{revtex4-1}

\usepackage{here}
\usepackage{comment,braket}
\usepackage{epsf}
\usepackage{amsmath}
\usepackage{graphics}
\usepackage{amsfonts}
\usepackage{dblfloatfix}
\usepackage{amssymb}
\usepackage{latexsym}
\usepackage{color}
\usepackage{ulem}
\input{colordvi.tex}
\usepackage{feynmp}
\usepackage[pdftex]{graphicx}
\usepackage{bm}
\usepackage[pdftex]{hyperref}
\usepackage[svgnames]{xcolor}
\definecolor{phthaloblue}{rgb}{0.0, 0.06, 0.54}
\hypersetup{
    colorlinks=true,
    linkcolor=phthaloblue,
    citecolor=blue,
    filecolor=blue,
    urlcolor=phthaloblue,
    }

\newcommand{\beq}{\begin{eqnarray}} 
\newcommand{\eeq}{\end{eqnarray}}

\def\({\left(}
\def\){\right)}
\def\[{\left[}
\def\]{\right]}

\newcommand{\bel}[1] {\begin{equation}\label{#1}}
\newcommand{\beal}[1] {\begin{eqnarray}\label{#1}}
\newcommand{\be}{\begin{equation}}
\newcommand{\ee}{\end{equation}}
\newcommand{\bea}{\begin{array}} 
\newcommand{\eea}{\end{array}}

\newcommand{\CL}{{\tt ${\mathcal C}$osmo${\mathcal L}$attice}~}
\newcommand{\CLns}{{\tt ${\mathcal C}$osmo${\mathcal L}$attice}}
\newcommand{\maqcd}{m_{a}^{\rm QCD}}

\newcommand{\eV}{ \ {\rm eV} }

\newcommand{\MeV}{\  {\rm MeV} }
\newcommand{\GeV}{\  {\rm GeV} }

\begin{document}

\title{(Non-)Perturbative Dynamics of a Light QCD Axion:\\Dark Matter and the Strong CP Problem
}

\author{Raymond T. Co}
\affiliation{Physics Department, Indiana University, Bloomington, IN 47405, USA}
\author{Taegyu Lee}
\affiliation{Physics Department, Indiana University, Bloomington, IN 47405, USA}
\author{Owen P. Leonard\vspace{1ex}}
\affiliation{Physics Department, Indiana University, Bloomington, IN 47405, USA}

\preprint{CETUP2025-008}

\begin{abstract}
Considerable theoretical efforts have gone into expanding the reach of the QCD axion beyond its canonical mass--decay-constant relation. The $Z_\mathcal{N}$ QCD axion model reduces the QCD axion mass naturally, by invoking a discrete $Z_\mathcal{N}$ symmetry through which the axion field is coupled to $\mathcal{N}$ copies of the Standard Model. Before the QCD phase transition at temperature $T_{\rm QCD}$, the $Z_\mathcal{N}$ potential has a minimum at misalignment angle $\theta=\pi$. At $T_{\rm QCD}$, $\theta =\pi$ becomes a maximum; the axion potential becomes exponentially suppressed and develops $\mathcal{N}$ minima---only one of which actually solves the strong CP problem. Before $T_{\rm QCD}$, $\theta$ relaxes towards $\pi$. After $T_{\rm QCD}$, the axion field starts from around the hilltop and may have sufficient kinetic energy to overcome the newly suppressed potential barriers. Such a field evolution leads to non-perturbative effects via the self-interactions near the hilltop, which can cause the exponential growth of fluctuations and backreaction on the coherent motion. This behavior can influence the relic density of the field and the minimum in which it settles. We conduct the first lattice simulations of the $Z_{\mathcal{N}}$ QCD axion using \CL to accurately calculate dark matter abundances and find non-perturbative dynamics reduce the abundance by up to a factor of two. We furthermore find that the probability of solving the strong CP problem tends to diverge considerably from the na\"\i ve expectation of $1/\mathcal{N}$.    
\end{abstract}

\maketitle

\section{Introduction}
\label{sec:Introduction}

Axions have emerged as compelling candidates in the search for dark matter~\cite{Preskill:1982cy, Dine:1982ah,Abbott:1982af}. In particular, the QCD axion attracts significant attention because it not only provides a promising dark matter candidate but is also an elegant solution to the strong CP problem~\cite{Peccei:1977hh, Peccei:1977ur,Weinberg:1977ma,Wilczek:1977pj}. The QCD axion is a pseudo Nambu-Goldstone boson which has its origins in the Peccei-Quinn (PQ) mechanism~\cite{Peccei:1977hh, Peccei:1977ur}. The PQ mechanism postulates an anomalous global symmetry, denoted by $U(1)_{\rm PQ}$, and a complex scalar field, $P$, that is charged under $U(1)_{\rm PQ}$. The angular mode of $P$, $\phi$ (the QCD axion), combines with the $\bar\theta$ parameter of the QCD Lagrangian, generating an effective $\bar\theta$ parameter, $\bar\theta_{\rm eff} \equiv \bar\theta+\phi/f_a$, where $f_a$ is the axion decay constant and we also define the misalignment angle $\theta \equiv \phi/f_a$. $U(1)_{\rm PQ}$ is spontaneously broken at the PQ phase transition when $P$ develops a radial vacuum expectation value. The PQ symmetry is then explicitly broken by QCD instanton effects at temperature $T_{\rm QCD} \simeq 150  \,\,\rm MeV$, generating a periodic potential for $\phi$, with a minimum corresponding to the CP-conserving $\bar\theta_{\rm eff}=0$\footnote{Henceforth, we implicitly employ a field redefinition to absorb the constant $\bar\theta$ and make the CP-conserving minimum correspond to $\theta=\phi/f_a=0$.}. As the universe cools, the axion begins to oscillate and eventually settle at this minimum, dynamically canceling the CP-violating term of the QCD Lagrangian.

Although compelling, the non-perturbative QCD effects which determine the axion potential result in a strict relation between $f_a$ and the canonical QCD axion mass $\maqcd$, which, for the chiral potential, takes the form $\maqcd \simeq 5.7 \mu{\rm eV} \times \left( 10^{12} {\rm GeV} / f_a \right)$~\cite{GrillidiCortona:2015jxo}. Current and upcoming axion searches mainly probe the parameter space for axions lighter than that predicted by this relation, where the axion interacts more strongly with the SM. If detection occurs in these regions, a few questions are of great interest. (i) Can such an axion be the QCD axion, which has a coupling to the gluons? (ii) If so, can this QCD axion actually solve the strong CP problem? (iii) Can it account for the observed dark matter abundance?

Many models have been proposed to relax the aforementioned canonical constraint on $m_a$~\cite{Dimopoulos:1979pp,Tye:1981zy,Holdom:1982ex,Flynn:1987rs,Rubakov:1997vp,Berezhiani:2000gh,Hook:2014cda,Fukuda:2015ana,Gherghetta:2016fhp, Dimopoulos:2016lvn, Agrawal:2017ksf, Agrawal:2017evu, Gaillard:2018xgk, Lillard:2018fdt, Fuentes-Martin:2019bue, Csaki:2019vte, Hook:2019qoh, Gherghetta:2020keg, Gherghetta:2020ofz, Valenti:2022tsc,Kivel:2022emq,Elahi:2023vhu,Co:2024bme}, where $m_a$ refers to a generic {\it QCD axion} mass not restricted by the canonical $m_a$--$f_a$ relation. The $Z_\mathcal{N}$ model, proposed by Hook~\cite{Hook:2018jle}, is a unique model that naturally allows for a lighter QCD axion. This model replicates the full Standard Model (SM) gauge and matter content $\mathcal{N}$ times and ties the copies together with a discrete $Z_{\mathcal{N}}$ symmetry acting on the axion. Each sector obtains an effective QCD phase angle $\theta_j \equiv  \theta + 2\pi j /\mathcal{N}$, where $j=0$ is identified as our sector and remaining $\mathcal{N}-1$ replicas act as dark sectors (DS). The relative phase offset of the $\mathcal{N}-1$ dark sectors creates destructive interference in the combined QCD axion chiral potential, naturally yielding an exponentially lighter axion that can solve the strong CP problem for odd $\mathcal{N}$~\cite{DiLuzio:2021pxd}. For even $\mathcal{N}$, the potential does not have a minimum at the CP-conserving angle.

\begin{figure}
    \centering
    \includegraphics[width=1.0\linewidth]{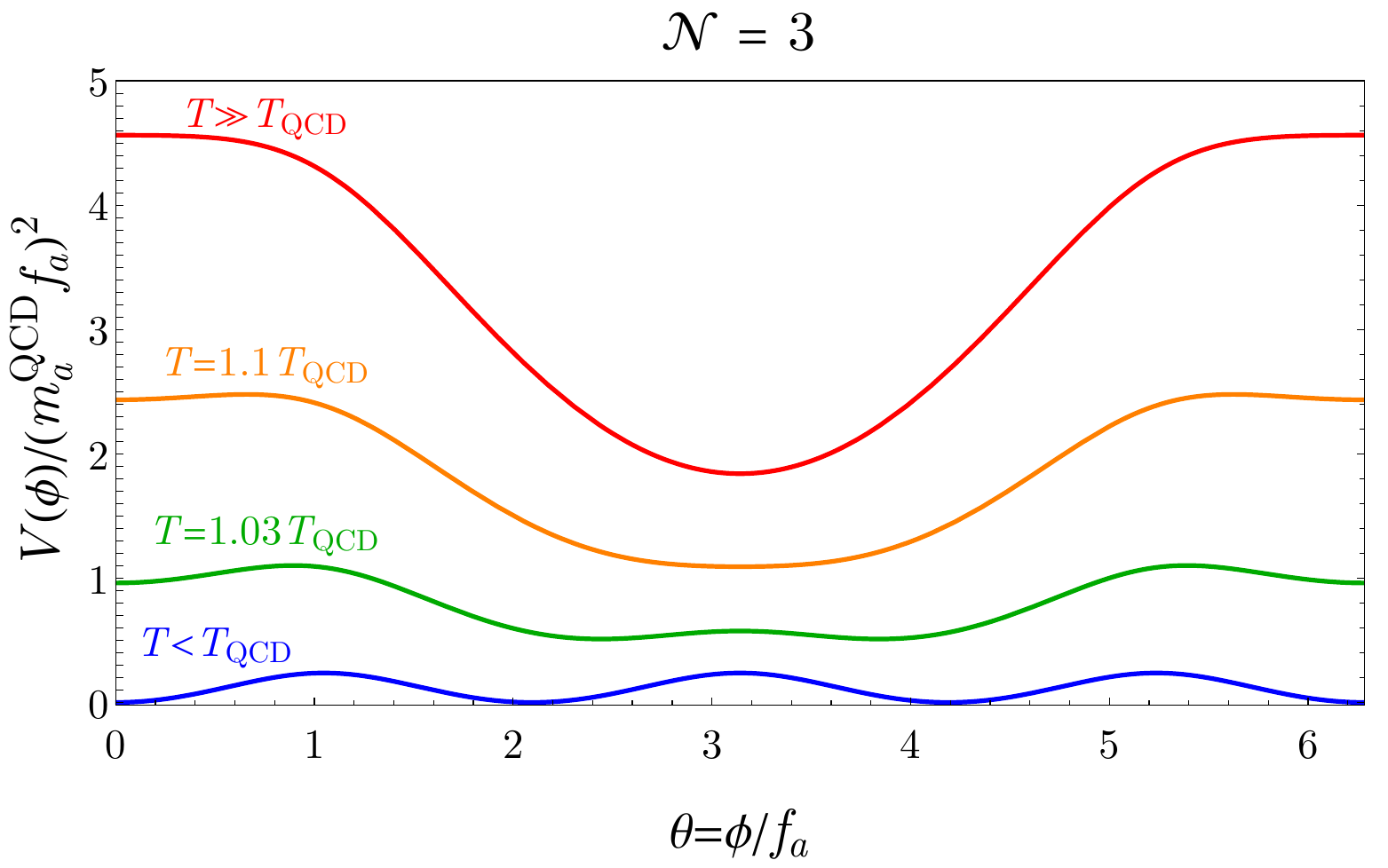}
    \caption{Temperature evolution of the $Z_{\mathcal{N}}$ potential for $\mathcal{N}=3$.}
    \label{fig:N3_Pot_Evolution}
\end{figure}

In a physical $Z_\mathcal{N}$ model, before the QCD phase transition the $Z_\mathcal{N}$ potential feels only the contributions from the DS, which conspire to create an unsuppressed potential with its minimum at $\theta=\pi$. At $T_{\rm QCD}$, the SM contribution fully turns on. All $\mathcal{N}$ terms of the potential sum in such a way to suppress and deform the potential. The deformation makes $\theta=\pi$ a maximum, and forms $\mathcal{N}$ new minima---including one at the CP-conserving misalignment angle $\theta=0$, see Fig.~\ref{fig:N3_Pot_Evolution}. The $Z_\mathcal{N}$ QCD axion initially has a mass comparable to $\maqcd$. Due to the suppression of the potential after $T_{\rm QCD}$, the mass becomes exponentially smaller than in the canonical one-sector case for the same decay constant. The extent of the decrease in mass is $\propto \mathcal{N}^{3/2}(m_u/m_d)^{\mathcal{N}}$ for large $\mathcal{N}$~\cite{DiLuzio:2021pxd} with $m_{u,d}$ as the up and down quark masses, broadening the domain of the QCD axion to the region $m_a < \maqcd \simeq 5.7 \mu{\rm eV} \times ( 10^{12} \GeV / f_a )$.

If the initial unsuppressed axion mass $\maqcd$ is larger than the Hubble rate at $T_{\rm QCD}$, $H(T_{\rm QCD})\equiv H_{\rm QCD}$, $\phi$ has a sizable nonzero kinetic energy from its initial relaxation to $\theta=\pi$. This kinetic energy is often greater than the potential energy of the suppressed potential barriers. This allows the field to enter a regime of rolling, a phenomenon known as kinetic misalignment~\cite{Co:2019jts}. When the field probes the hilltop it is affected nonlinearly by self-interactions via the quartic and higher order terms of its potential~\cite{Jaeckel:2016qjp,Berges:2019dgr,Fonseca:2019ypl}. Properties of the field, particularly its relic density and the minimum in which it settles, can be affected by these processes.

Great work has previously been done in considering the $Z_\mathcal{N}$ QCD axion as a dark matter candidate using perturbative numerical methods in Ref.~\cite{DiLuzio:2021gos}, and as a solution to the strong CP problem using analytical and perturbative numerical methods in Refs.~\cite{DiLuzio:2021pxd, DiLuzio:2021gos}. In the present work, we improve upon these works by leveraging lattice simulations of the $Z_\mathcal{N}$ model using the \CL ($\mathcal{CL}$) framework~\cite{Figueroa:2021yhd}. We simulate the evolution of the $Z_\mathcal{N}$ axion field in an expanding universe during radiation domination through the QCD phase transition.

In this work, we conduct the first lattice simulations of the $Z_\mathcal{N}$ model and identify the region where the phenomenology is significantly influenced by non-perturbative processes. We will review the $Z_\mathcal{N}$ model in Sec.~\ref{sec:ZN_Model} and analytically examine the equations of motion in Sec.~\ref{sec:Axion_Field_Evolution}. We perform a perturbative numerical analysis and lattice simulations in Sec.~\ref{sec:Analyses}. Specifically, we present an understanding of the parameters for which the $Z_\mathcal{N}$ QCD axion can provide a solution to the strong CP problem in Sec.~\ref{subsec:strongCP}.  After the QCD phase transition, the potential develops $\mathcal N$ degenerate minima, and it has been assumed that the axion field ends up in the CP-conserving minimum ($\theta=0$) with $1/\mathcal N$ probability\footnote{However, a model was recently proposed in which a soft $Z_{\mathcal N}$-breaking coupling between the SM Higgs and a reheaton lifts the $N$-fold vacuum degeneracy, selecting the $\theta=0$ minimum and eliminating the usual $1/\mathcal N$ tuning~\cite{Banerjee:2025zcd}. This fix, though, relies on postulating an explicit symmetry-breaking term of the right sign and size.}~\cite{Hook:2018jle,DiLuzio:2021gos}. We re-examine the $1/\mathcal N$ expectation in the perturbative regime by explicitly averaging over the distribution of initial misalignment angles and uncertainty in the QCD phase transition temperature. In Sec.~\ref{subsec:DM}, we present our results for the dark matter abundance. We further quantify the extent to which analytical and perturbative numerical predictions deviate from those obtained through lattice simulations, highlighting how non-perturbative dynamics---often neglected in axion cosmology---can significantly impact the phenomenology of these models. We summarize and discuss in Sec.~\ref{sec:summary}.

\section{The $Z_\mathcal{N}$ QCD Axion Model}
\label{sec:ZN_Model}
What follows is a brief overview of the vital aspects of the $Z_\mathcal{N}$ model, initially proposed by Hook~\cite{Hook:2018jle}, further developed analytically in Ref.~\cite{DiLuzio:2021pxd}, and considered as a dark matter candidate in the so-called trapped misalignment mechanism using semi-analytical and perturbative numerical methods in Ref.~\cite{DiLuzio:2021gos}.

As a starting point, consider the chiral potential of the canonical QCD axion, 
\begin{equation}
    V_{\rm SM}(\theta)=-m_{\pi}^2f_{\pi}^2\sqrt{1-\frac{4 m_u m_d}{(m_u+m_d)^2}\sin^2\left(\frac{\theta}{2}\right)} \, ,
    \label{eq:SM_Pot_Tindep}
\end{equation}
where $\theta = \phi/f_a$.

The $Z_\mathcal{N}$ model considers a QCD axion with couplings to $\mathcal{N}$ SM copies of the form
\begin{equation}
    \mathcal{L} \supset \frac{\alpha_s}{8\pi} \sum_{j=0}^{\mathcal{N}-1} \left(\theta+\frac{2 \pi j}{\mathcal{N}}\right)G_j\tilde{G}_j\;.
    \label{eq:N_SM_Couplings}
\end{equation}
Under these couplings the chiral potential contribution from the DS is
\begin{equation}
    V_{\rm DS}(\theta)=-m_{\pi}^2f_{\pi}^2\sum_{j=1}^{\mathcal{N}-1}\sqrt{1-\frac{4 m_u m_d}{(m_u+m_d)^2}\sin^2\left(\frac{\theta}{2}+\frac{ j \pi }{\mathcal{N}}\right)} \, .
    \label{eq:DS_Pot_Tindep}
\end{equation}

The chiral potential, Eq.~\eqref{eq:SM_Pot_Tindep}, is suppressed at temperatures greater than $T_{\rm QCD}$. We emulate this effect with the inclusion of a temperature-dependent coefficient $h(T)$~\cite{DiLuzio:2021gos}, which introduces a power law suppression to Eq.~\eqref{eq:SM_Pot_Tindep} for temperature $T>T_{\rm QCD}$
\begin{equation}
    V_{\rm SM}(\theta,T)=h(T)\times V_{\rm SM}(\theta),
    \label{eq:SM_Pot_Tdep}
\end{equation}
where
\begin{equation}
    h(T)\equiv \begin{cases}
        \left(\frac{T_{\rm QCD}}{T}\right)^\beta \qquad T>T_{\rm QCD} \\
        \quad \,1 \qquad \qquad T<T_{\rm QCD} 
    \end{cases}\,.
    \label{eq:Pot_Powerlaw_Tdep}
\end{equation}
In this work, we use an $\beta=8$ obtained from dilute instanton gas approximations~\cite{Pisarski:1980md}, which is in agreement with lattice QCD simulations done in Refs.~\cite{Berkowitz:2015aua,Kitano:2015fla,Borsanyi:2015cka,Petreczky:2016vrs,Borsanyi:2016ksw,Taniguchi:2016tjc}. However, we will comment on the extent to which $h(T)$ affects the dark matter abundance.

A physically valid $Z_{\mathcal{N}}$ model requires that we take all DS to be sufficiently cold and well approximated by their zero-temperature potentials by the time the temperature dependence of the SM becomes relevant. Hence, we can combine Eqs.~\eqref{eq:DS_Pot_Tindep} and \eqref{eq:SM_Pot_Tdep} to get the full potential\footnote{Throughout our analysis, we eliminate the zero-temperature constant vacuum energy of the potential.}
\begin{align}
V_\mathcal{N}(\theta,T)=V_{\rm SM}(\theta,T) + V_{\rm DS}(\theta).
\label{eq:ZN_Pot}
\end{align} 

It has been shown in Ref.~\cite{DiLuzio:2021pxd} that, in the large $\mathcal{N}$ limit, the zero-temperature $Z_\mathcal{N}$ potential, given in Eq.~\eqref{eq:ZN_Pot}, is well approximated by
\begin{equation}
    V_\mathcal{N}(\theta,T=0) \simeq -\frac{m_a^2 f_a^2}{\mathcal{N}^2}\cos\left(\frac{\mathcal{N}\phi}{f_a}\right).
    \label{eq:simp_ZNpot}
\end{equation}
In this same limit, Ref.~\cite{DiLuzio:2021pxd} has also shown that, after the $Z_\mathcal{N}$ potential is fully turned on, the dependence of the suppressed mass $m_a$ on $\mathcal{N}$ is well approximated by
\begin{equation}
    m^2_a f^2_a \simeq \frac{m^2_\pi f^2_\pi}{\sqrt{\pi}}\sqrt{\frac{1 - z}{1 + z}}\mathcal{N}^{3/2}z^{\mathcal{N}} ,
\end{equation}
where $z\equiv m_u/m_d \simeq 0.48$~\cite{GrillidiCortona:2015jxo}. Hence, the mass experiences an exponential suppression 
\begin{equation}
    \left( \frac{m_a}{\maqcd} \right)^2=  \mathcal{N}^{3/2} z^{\mathcal{N}-1}(1+z)^2 \sqrt{\frac{1-z}{\pi (1+z)}}.
    \label{eq:supp}
\end{equation}

\section{Axion field evolution}
\label{sec:Axion_Field_Evolution}
Since the effective potential has $\mathcal{N}$ degenerate minima, if the PQ symmetry is spontaneously broken after inflation, the explicit breaking at $T_{\rm QCD}$ creates stable domain walls between these minima, resulting in an unacceptable domination of the cosmic energy density today. Therefore, the scenario in which the $Z_\mathcal{N}$ model is physically viable is when spontaneous PQ breaking occurs pre-inflation, which we consider in this work. To satisfy the current bounds on the exotic effective relativistic degrees of freedom $\Delta N_{\rm eff}\leq0.3$ \cite{Planck:2019nip}, we require DS radiation to have a sufficiently small energy density, e.g., by asymmetric reheating, and hence the DS temperatures are much lower than the SM one.

These conditions necessitate that the DS contributions to the potential, as seen in Eq.~\eqref{eq:DS_Pot_Tindep}, are fully turned on well before $T_{\rm QCD}$. Only the SM contribution, given in Eq.~\eqref{eq:SM_Pot_Tdep}, has a temperature dependence relevant to this stage. Thus, before $T_{\rm QCD}$, $V_\mathcal{N}\simeq V_{\rm DS}$. Moreover, $V_{\rm DS}\simeq -V_{\rm SM}$~\cite{DiLuzio:2021gos}, and thus the potential initially has a minimum at $\theta=\pi$, and the axion mass approximately obeys the canonical $m_a^{\rm QCD}$-$f_a$ relation\footnote{Before the QCD phase transition, the mass of the axion near $\theta\simeq\pi$ is not exactly $m_a^{\rm QCD}$ but $\maqcd\sqrt{(1+z)/(1-z)}$, whose difference is roughly $1.7$, with an additional $\mathcal{O}(1)$ dependence on the field value. In the qualitative and analytic discussions, we neglect this factor for simplicity, whereas all numerical and lattice results employ the full potential and therefore include them.}. For $T \le T_{\rm QCD}$, the $Z_{\mathcal N}$ potential is fully turned on and is time-independent, $V_\mathcal{N}(\theta) = V_{\rm DS} + V_{\rm SM}$. The DS and SM contributions experience a near-precise cancellation, resulting in the exponential suppression of the potential which drives the corresponding reduction in $m_a$. In addition, the single minimum at $\theta= \pi$ becomes $\mathcal{N}$ degenerate minima, including the CP-conserving minimum at $\theta =0$. The stages of the evolution are depicted graphically in Fig.~\ref{fig:N3_Pot_Evolution} for $\mathcal{N} = 3$.

In an expanding Friedmann-Lemaître-Robertson-Walker metric, the time evolution of the axion field is governed by the equation of motion (EOM)
\begin{equation}
\ddot{\phi}+3H\dot{\phi}-\frac{1}{a^2}\nabla^2\phi+V'(\phi)=0
    \label{eq:Damped_KG_EQ_Full} ,
\end{equation}
with $V'(\phi) \equiv d V(\phi) / d \phi$, and dots refer to derivatives with respect to cosmic time.
Neglecting any interactions between the spatially homogeneous zero mode $\phi_0$ and the fluctuations around the zero mode $\delta \phi=\phi-\phi_0$, the EOM of the zero mode is
\begin{equation}
    \ddot{\phi}_0+3H\dot{\phi}_0+V'(\phi_0)=0,
    \label{eq:zeromode_EOM}
\end{equation}
where $H$ is the Hubble rate. We can decompose $\delta\phi$ into its individual Fourier modes $\delta\phi_k$. The relation between $\delta\phi$ and $\delta\phi_k$ is given explicitly by its Fourier transform,
\begin{equation}
    \delta\phi=\int \frac{d^3k}{8\pi^3}\delta\phi_ke^{-i{\bf k} \cdot {\bf x}}.
    \label{eq:FT_deltaphi_deltaphik}
\end{equation}
Assuming that the $k$-modes are also non-interacting, the EOM of $\delta\phi_k$ is
\begin{equation}
    \delta \ddot{\phi}_k+3H\delta\dot{\phi}_k+\left[\left(\frac{k}{a}\right)^2+V''(\phi_0)\right]\delta\phi_k=0,
    \label{eq:fluct_EOM}
\end{equation}
where $a$ is the scale factor.

A detailed analytic description for the evolution of the $k$-modes governed by Eq.~\eqref{eq:fluct_EOM} requires Floquet analysis. In the kinetic misalignment scenario, such an analysis has been done very thoroughly in Ref.~\cite{Eroncel:2022vjg}. Similarly this analysis has been done in the large amplitude oscillation scenario in Ref.~\cite{Arvanitaki:2019rax}. The most important phenomenon is parametric resonance (PR)~\cite{Kofman:1994rk,Kofman:1997yn}, which causes the exponential growth of certain $\delta\phi_k$. For the sake of this work, we provide a brief qualitative description of the conditions under which PR arises and the effect it has on the field. PR typically occurs for a particular $k$-mode when the oscillation frequency of $V''(\phi_0)$ resonates with the mode's natural frequency, $\omega_{k,0}=k/a$. The oscillation frequency of $V''(\phi_0)$ is often denoted as the pump frequency,  $\omega_{\rm pump}$, because the zero-mode oscillations ``pump" energy into the fluctuations. In such a situation $\delta\phi_k$ grows exponentially. Resonance occurs only when the parameter $V''(\phi_0)$ is time-varying, such that the zero mode can pump $\delta\phi_k$. In the $Z_\mathcal{N}$ model, this can occur frequently when $\phi_0$ probes the anharmonic regions of the potential either via oscillations after trapping or kinetic misalignment before trapping. Since PR underpins the emergence of non-perturbative dynamics in the field evolution, deriving analytic criteria to indicate which $\delta \phi_k$ resonate with $\phi_0$ in both regimes will provide a substantial benefit in interpreting our results. In this section, for simplicity, we will use the greatly simplified large $\mathcal{N}$ potential, given in Eq.~\eqref{eq:simp_ZNpot}. Note that only PR due to large amplitude oscillation takes place in the parameter space considered in this work. However, both scenarios generically arise in the $Z_{\mathcal{N}}$ model.

In the case of PR due to large amplitude oscillations after trapping, $\omega_{\rm pump}$ is around $2m_a$. To see this, consider the curvature of Eq.~\eqref{eq:simp_ZNpot}. Assuming that the period of oscillation is much less than Hubble time, and since $\phi$ spends most of an oscillation in the harmonic regime of its potential, we can approximate its motion using the parametrization $\phi \approx \Phi \cos(m_a t)$ with $\Phi$ the amplitude. Using this parameterization the curvature takes the form
\begin{equation}
    V_{\mathcal{N}}''(t) \approx m_a^2\cos\left( \frac{\mathcal N \Phi \cos(m_a t)}{f_a} \right).
    \label{eq:curvature_parameterization}
\end{equation}
Expressing the outermost cosine in exponential form, we can rewrite Eq.~\eqref{eq:curvature_parameterization} using a Jacobi-Anger expansion,
\begin{equation}
    V_{\mathcal{N}}''(t)
    \approx m_a^2
    \bigg[
      J_0\!\left(x\right)
      + 2 \sum_{j=1}^{\infty} (-1)^j\,
        J_{2j}\!\left(x\right)
        \cos\!\left(2j m_a t\right)
    \bigg] ,
\end{equation}
where $x = \mathcal{N}\Phi/f_a$ and the $J_\alpha(x)$ are Bessel functions of the first kind. For $x<\pi$, the sum is dominated by $J_2(x)$. Truncating after the first oscillatory term, the curvature takes the form
\begin{equation}
    V_{\mathcal{N}}''(t) \approx m_a^2 \left[ J_0 \left( x \right) - 2 J_{2}\left( x \right)\cos {(2  m_a t)}    \right].
    \label{eq:oscillatory_term}
\end{equation}
Hence, we can read off that $\omega_{\rm pump}=~2m_a$. 

Since we observe PR resulting from large oscillations after trapping in this work, we can take a step further to determine the dominant instability band of the resonance. First we can write the Mathieu equation in order to find the instability band,
\begin{equation}
    \frac{d^2}{d\tilde{t}^2}\delta\phi_k + \left(A_k -2q \cos\left(2\tilde{t}\right)\right)\delta\phi_k =0.
    \label{eq:mathieu}
\end{equation}
Using Eq.~\eqref{eq:fluct_EOM}, Eq.~\eqref{eq:oscillatory_term}, change of variables to dimensionless time, $\tilde{t} = m_a t$, and further assuming Hubble scale is negligible compared to $m_a$, we identify the parameters in Eq.~\eqref{eq:mathieu} to be
\begin{equation}
      A_k=\left(\frac{k}{am_a}\right)^2 + J_0 \left( x \right), \qquad q=J_{2}\left( x \right).
\end{equation}
The Mathieu equation has an instability band around $A_k\sim n^2$ where $n=1,2,3,\cdots$. For $q\ll1$, modes that fall inside these bands grow exponentially, $e^{\mu_k \tilde t}$, with a characteristic Floquet exponent $\mu_k \simeq q/2=J_2(x)/2$. Thus, the mode that experiences the largest growth occurs for $x\simeq 3.05$, indicating that large amplitude oscillations are necessary. For $x\simeq 3.05$, $q\simeq 0.49$, which could modify the Floquet exponent slightly. The mode experiencing the largest growth due to the instability is $k/a \approx m_a \sqrt{1-J_0\left(x\simeq 3.05\right)} \simeq 1.13m_a$. This has been confirmed for some benchmark points during the initial stage of PR on the lattice.

For PR due to kinetic misalignment, recall that the evolution of $\phi$ in this regime is characterized by rolling over potential barriers. Assuming that $\phi$ has enough kinetic energy such that the time it takes to roll through one period is much less than the Hubble time and that $\dot\phi$ is approximately constant over a period, we can use the parameterization $\phi=\dot\phi t$. Under this parameterization the curvature becomes
\begin{equation}
    V''_\mathcal{N}(t)\approx m_a^2\cos\left(\frac{\mathcal{N}\dot\phi}{f_a}t\right).
\end{equation}
From this we can read off that $\omega_{\rm pump}={\mathcal{N}\dot\phi}/{f_a}$, during kinetic misalignment.

The longer $\phi$ probes the anharmonic regimes of its potential, be it through rolling or large amplitude oscillations, the longer PR occurs. In an expanding universe, then, the smaller $m_a$ is relative to $3H$ after trapping, the more efficient PR will be. We now outline the various components of our analysis. For a concise overview of the region of the parameter space this work concerns, as well as the numerical values for the physical parameters we used in our calculations, see App.~\ref{sec:appendix_A}. This work is two-pronged: an exhaustive zero-mode analysis and targeted lattice simulations. In both cases the general goal is to determine the parameter space where the $Z_\mathcal{N}$ QCD axion can account for the observed dark matter abundance and solve the strong CP problem by the field settling to the minimum at zero. 

\section{Analyses}
\label{sec:Analyses}

We now present analyses on the strong CP problem as well as the dark matter abundance. In both cases, we begin with considering the zero-mode evolution by neglecting the fluctuations, an approach that will be shown to be invalid in certain regions of the parameter space. We follow this by the full field evolution on the lattice.
Although the zero-mode approach is not valid everywhere, it is tremendously helpful to develop some intuitive understanding by studying the dynamics throughout the viable parameter space in great detail. The two main parameters that govern the dynamics of the model are 1) the initial unsuppressed mass $\maqcd$ and 2) the amount of mass suppression, or equivalently, $\mathcal{N}$.

Based on Eq.~\eqref{eq:zeromode_EOM}, if $\maqcd \gtrsim 3H_{\rm QCD} \simeq 4 \times 10^{-11} \eV$, the field will begin relaxing towards $\theta=\pi$ before $T_{\rm QCD}$. For a larger initial unsuppressed mass, oscillations begin earlier, so the field experiences a longer period of redshifting. In this case, its kinetic energy is partially depleted due to redshift, and the misalignment angle has relaxed near $\theta=\pi$. For a smaller initial unsuppressed mass, the redshift of the field amplitude starts much closer to $T_{\rm QCD}$, and the field loses less of its kinetic energy. As a result, a QCD axion with larger $\maqcd$, or equivalently smaller $f_a$, cannot solve the strong CP problem because the field does not have sufficient kinetic energy to roll over the $(\mathcal{N}-1)/2$ barriers necessary to end up in the CP-conserving minimum of the degenerate potential.

For a given $\maqcd$, $\mathcal{N}$ dictates the amount of the mass suppression given in Eq.~\eqref{eq:supp}. For large $\mathcal{N}$, the potential after $T=T_{\rm QCD}$ is much more suppressed, which enables the axion to more easily roll over a sufficient number of barriers to the CP-conserving minimum. As a result, larger $\mathcal{N}$ helps with solving the strong CP problem.

In addition to the model parameters $\maqcd$ and $\mathcal{N}$, the exact timing of the QCD phase transition $T_{\rm QCD}$ and the initial misalignment angle $\theta_i$ at $T \gg T_{\rm QCD}$ affect the dynamics in non-trivial ways. We now discuss these effects in order.

The temperature $T_{\rm QCD}$ that determines the axion mass scaling in Eq.~\eqref{eq:Pot_Powerlaw_Tdep} is close to the QCD crossover temperature around $150 \MeV$. Fitting to lattice QCD results, Ref.~\cite{Hoof:2021jft} finds $T_{\rm QCD} \simeq 143.7 \pm 2.9 \MeV$, which corresponds to a 2\% uncertainty. We take the power-law form of Eq.~\eqref{eq:Pot_Powerlaw_Tdep} and $T_{\rm QCD} = 143.7 \MeV$. However, in our zero-mode analyses, we allow $T_{\rm QCD}$ to vary by 2\% based on the uncertainty obtained in Ref.~\cite{Hoof:2021jft}. The exact timing of $T_{\rm QCD}$ in relation to the phase of an axion oscillation cycle determines the axion field velocity and therefore can affect significantly which final minimum the axion falls into. When the oscillation period, $2\pi/\maqcd$, is much smaller than the uncertainty in the QCD timing $\Delta t_{\rm QCD}$, we simply vary the timing over one oscillation cycle since varying over additional cycles will simply produce similar results. In the opposite case where $2\pi/\maqcd \gtrsim \Delta t_{\rm QCD}$, we vary $T_{\rm QCD}$ by its full 2\% uncertainty.

\subsection{Strong CP problem}
\label{subsec:strongCP}
\begin{figure}
    \centering
    \includegraphics[width=1.0\linewidth]{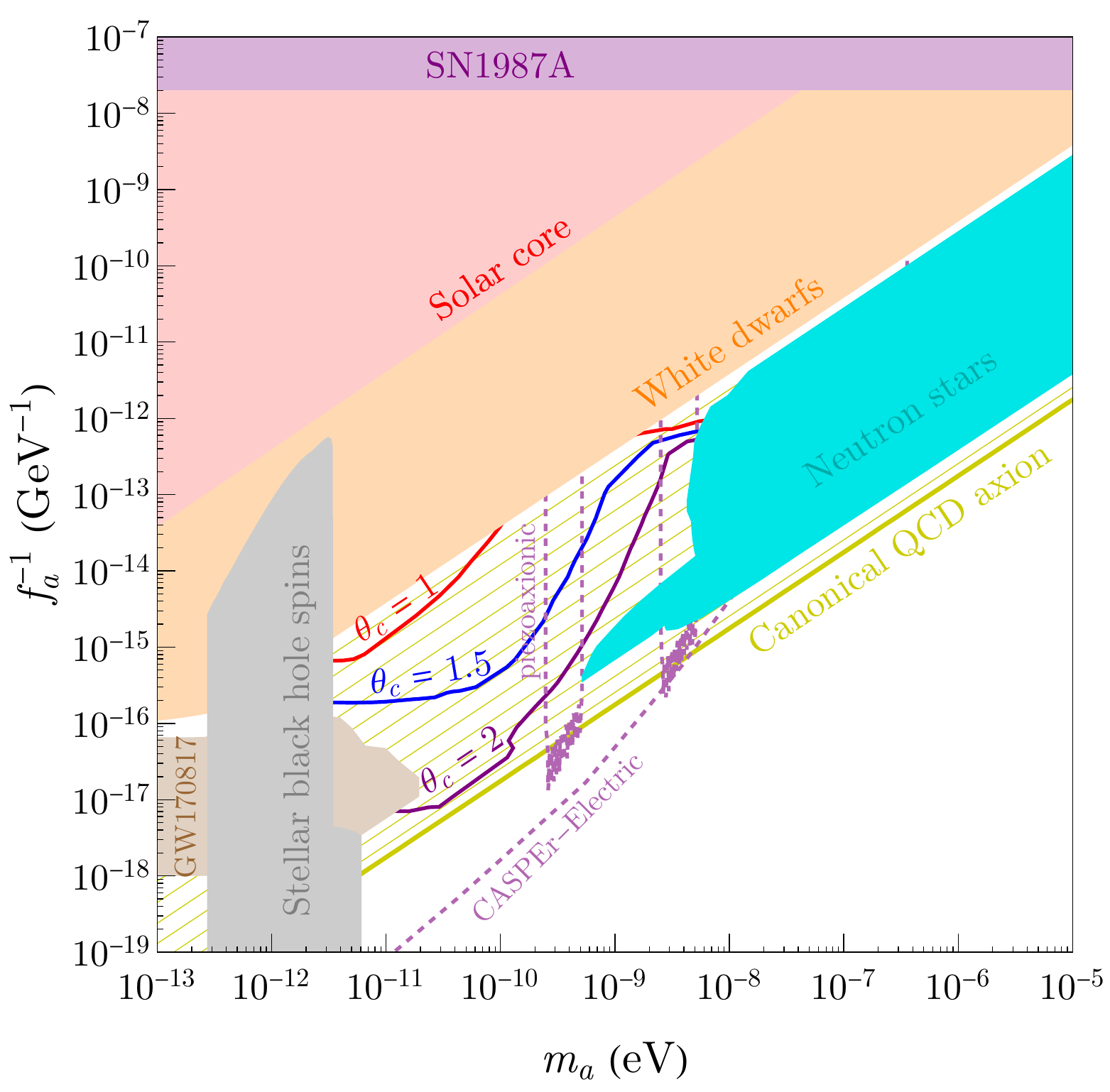}
    \caption{Contours of the critical amplitude $\theta_c$, which is the minimum initial amplitude $|\pi-\theta_i|$ necessary for the axion to reach the CP-conserving minimum after the QCD phase transition. The thin yellow lines show different values of $Z_{\mathcal{N}}$ with $Z_{\mathcal{N}}=27$ near the white dwarfs constraint and $Z_{\mathcal{N}}=5$ above the canonical QCD axion line, whereas $Z_{\mathcal{N}}=3$ is too close to the QCD axion line to be visible. The shaded regions show astrophysical constraints from supernovae~\cite{Springmann:2024ret}, the solar core~\cite{Hook:2017psm}, white dwarf composition~\cite{Balkin:2022qer}, neutron star cooling~\cite{Gomez-Banon:2024oux,Kumamoto:2024wjd}, stellar black hole spins~\cite{Baryakhtar:2020gao}, and axion-nucleon couplings from GW170817~\cite{Zhang:2021mks}. Regions above the purple dashed lines are within the reach of CASPEr-Electric~\cite{JacksonKimball:2017elr} and the experiment involving the piezoaxionic effect~\cite{Arvanitaki:2021wjk}.}
    \label{fig:theta_c}
\end{figure}

The initial amplitude $|\pi - \theta_i|$ directly determines the initial potential energy, and consequently the residual kinetic energy at $T_{\rm QCD}$, thereby affecting the ability of the axion misalignment angle $\theta$ to roll to an integer multiple of $2\pi$. This implies that, for every point in the parameter space, there is a critical value $\theta_c$, above which $|\pi -\theta_i|$ is required for the axion to solve the strong CP problem. The value of $\theta_c$ is determined as follows, using the zero-mode approach. For each initial amplitude, we marginalize over the uncertainty of $T_{\rm QCD}$ using 256 sample points with the algorithm described in the previous paragraph. If the initial amplitude is above (below) $\theta_c$, the axion does (not) fall into the CP-conserving minimum in at least one (any) of these 256 points. The resultant contours of the critical amplitude $\theta_c$ are shown in Fig.~\ref{fig:theta_c}. This crucial criterion for solving the strong CP problem has not been pointed out previously in the literature. 

\begin{figure}
    \centering
    \includegraphics[width=1.0\linewidth]{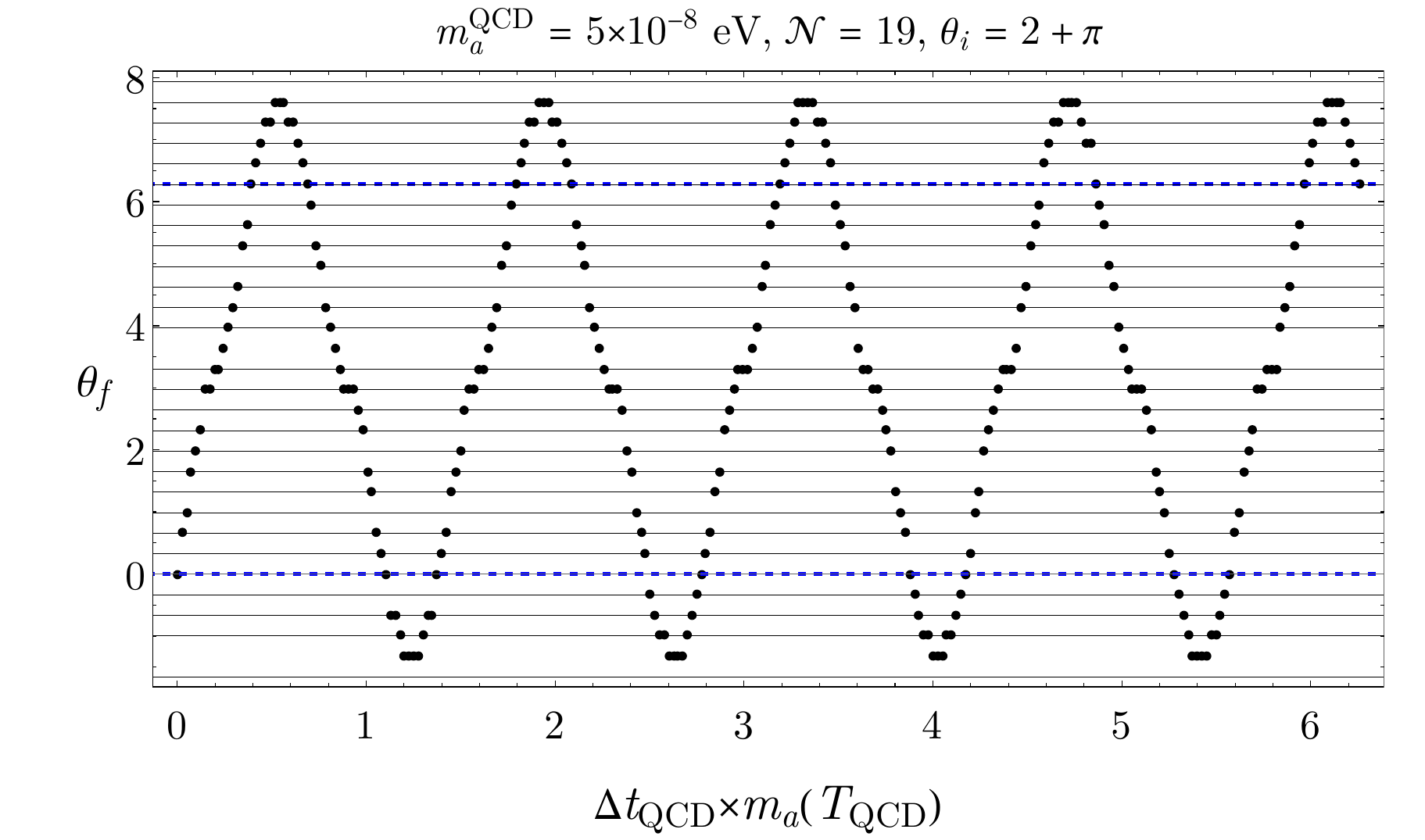}
    \includegraphics[width=1.0\linewidth]{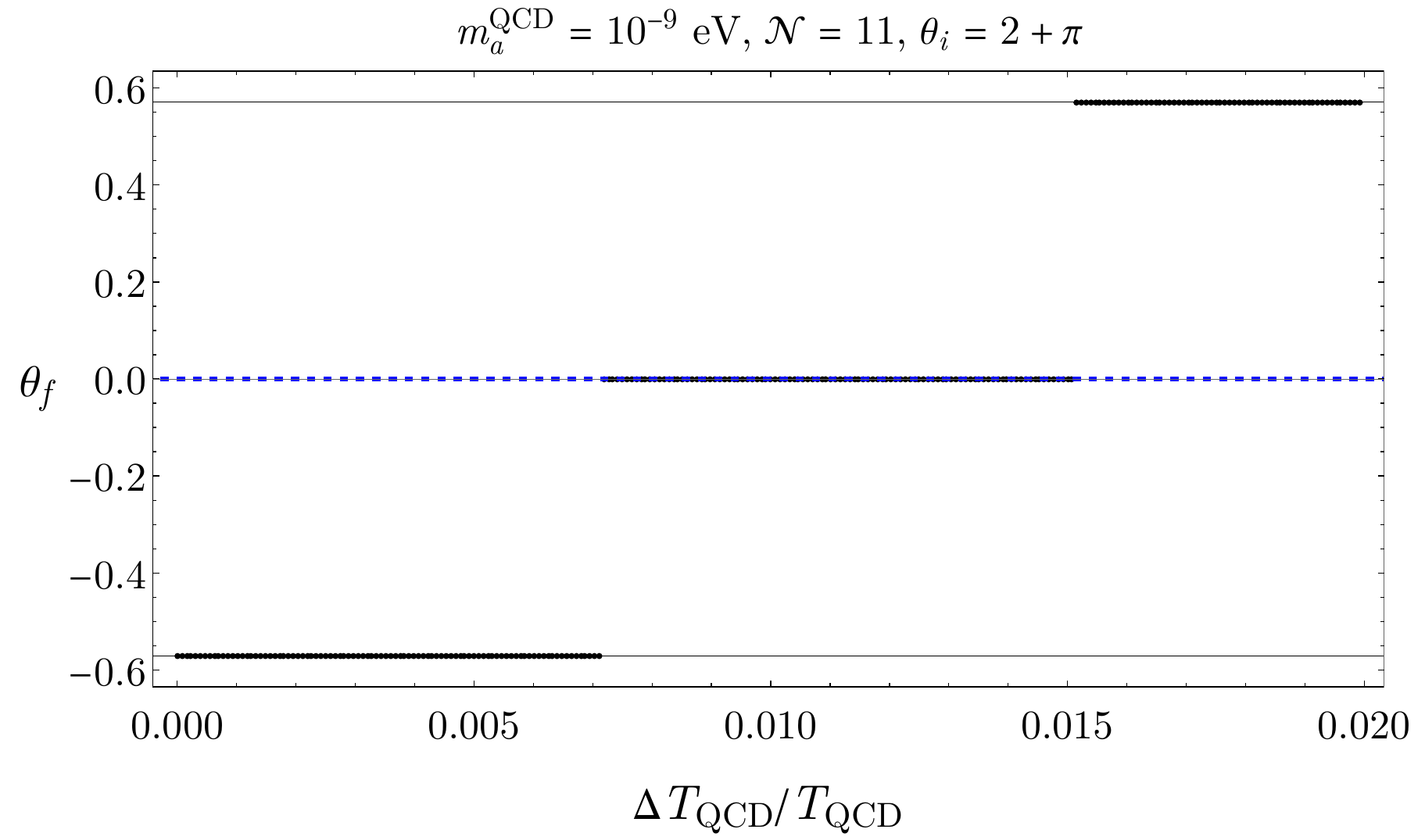}
    \caption{The black dots show the final minima that the axion settles in as a function of the QCD time/temperature variation (upper/lower panel) from the zero-mode analysis with parameters specified at the top axes. The horizontal gray lines indicate the location of the potential minima, while the CP-conserving minimum is at the blue dashed lines.}
    \label{fig:final_min}
\end{figure}

For a large mass, we indeed observe the periodic features in the final minimum as alluded to earlier. This is demonstrated in the upper panel of Fig.~\ref{fig:final_min}. For the low mass regime, we note that the axion field barely starts to roll before $T_{\rm QCD}$, and the full 2\% variation has a limited effect on the final minimum reached as shown in the lower panel of Fig.~\ref{fig:final_min}.

The computational cost required to recreate such a sampling on the lattice makes the endeavor unfeasible for this work. Instead, we initially simulate each benchmark point once, for two misalignment angles $\vert \pi - \theta_i \vert = \{1,\,2\}$. In addition, we examine the effect of the timing of $T_{\rm QCD}$ with the same method as in the zero-mode analysis, varying over 13 points instead of 256, for selected benchmark points, which are specified in App.~\ref{sec:appendix_B}. From the lattice results, we find that the energy content of $\delta \phi_k$ remains subdominant before trapping in the region of the parameter space considered in this work. As a result, we expect any non-perturbative dynamics will have negligible effects on $\theta_c$. This expectation was confirmed for $\vert \pi - \theta_i \vert = 2$ by conducting the 13-point phase variations mentioned above for benchmark points along the $\theta_c$ boundary determined in the zero-mode analysis.

\begin{figure}
    \centering
    \includegraphics[width=1.0\linewidth]{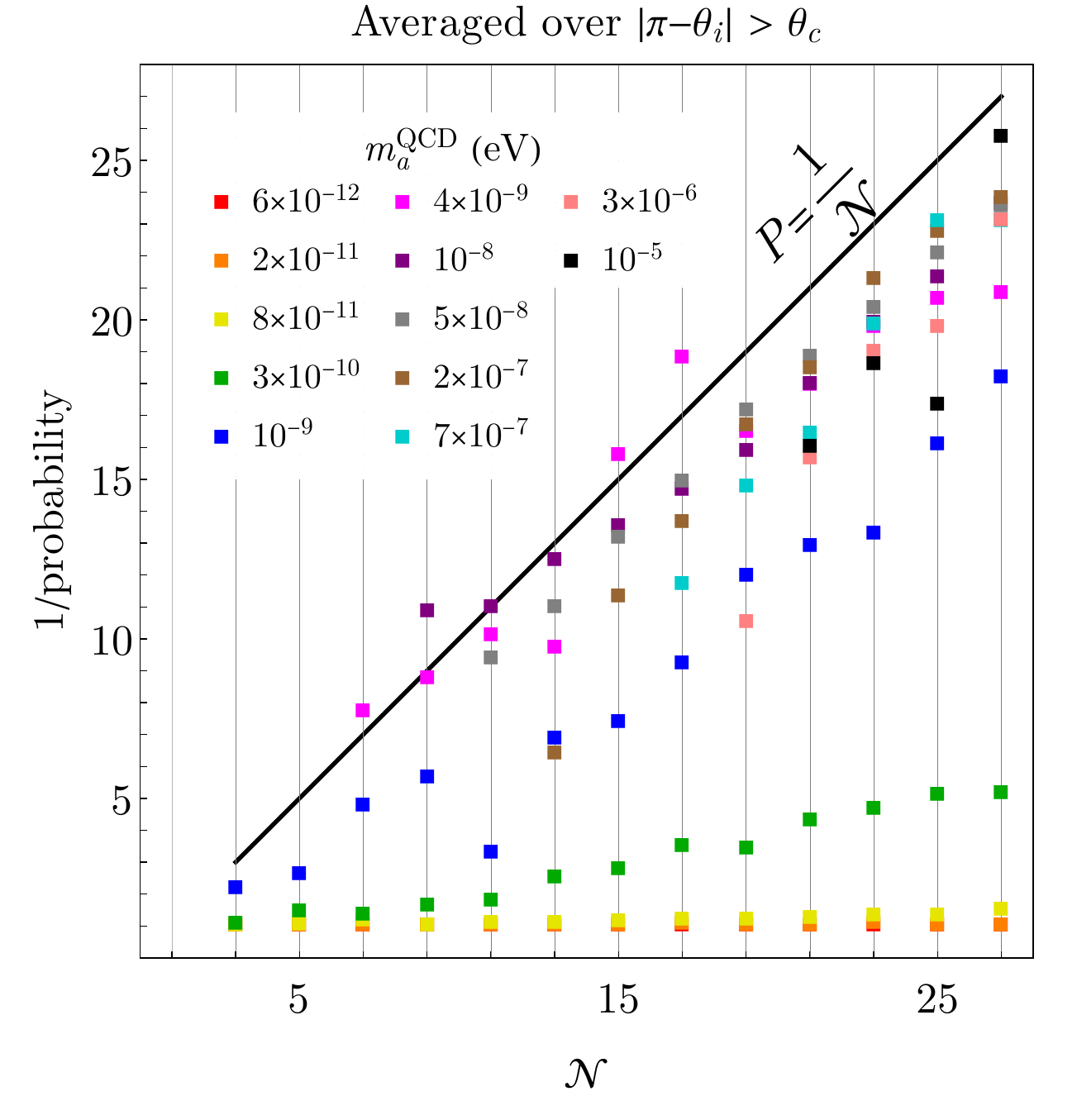}
    \caption{The average probability of solving the strong CP problem, accounting for variations in the initial misalignment angle $\theta_i$ and the uncertainty in the QCD phase transition temperature $T_{\rm QCD}$. The timing of the QCD phase transition is varied up to an axion oscillation cycle or over the theoretical 2\% uncertainty in $T_{\rm QCD}$, whichever is smaller. Here, the initial amplitude $|\pi-\theta_i|$ is restricted to values above the critical angle $\theta_c$, below which the probability vanishes identically. The solid black line indicates the na\"\i ve expectation,~$P=1/\mathcal{N}$.}
    \label{fig:prob1}
\end{figure}

\begin{figure}
    \centering
    \includegraphics[width=1.0\linewidth]{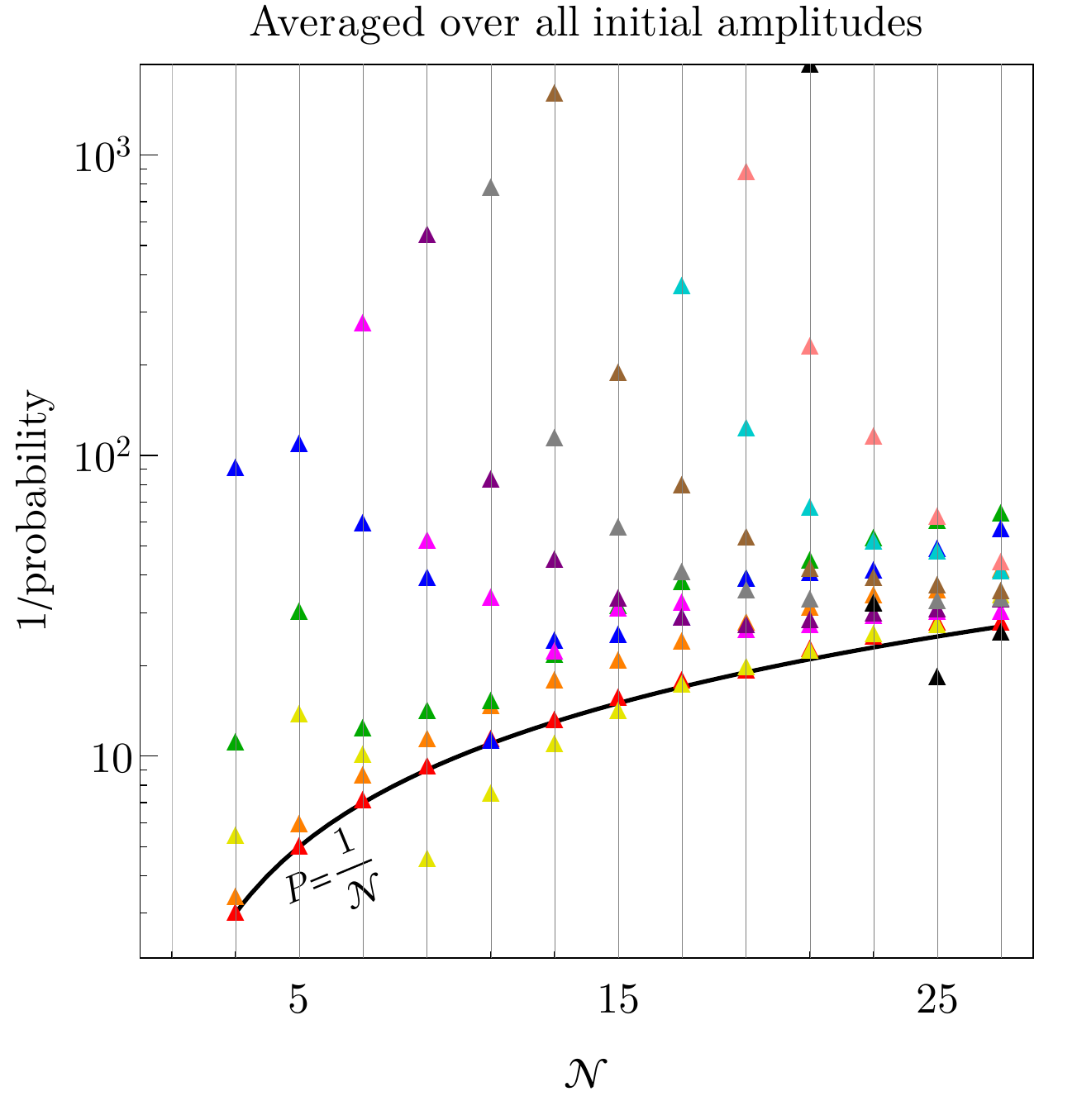}
    \caption{Same as Fig.~\ref{fig:prob1}, but with $|\pi - \theta_i|$ allowed to vary over the full range $(0, \pi)$, now including the ranges of $\theta_i$ where probability vanishes. The $y$-axis is on logarithmic scale due to the substantially lower probabilities.}
    \label{fig:prob_overall}
\end{figure}

We now examine, within the zero-mode approximation, the probability that the $Z_\mathcal{N}$ QCD axion can solve the strong CP problem. The lattice results summarized above, in the discussion of $\theta_c$, indicate that non-perturbative dynamics remain subdominant \textit{before trapping} across the parameter space considered in this work. As a result, the probability obtained from the zero-mode analysis should provide a reliable estimate even in the non-perturbative regime. Previous literature only provided a general hypothesis that the probability of ending in the CP-conserving minimum is $1/\mathcal{N}$~\cite{Hook:2018jle,DiLuzio:2021gos}. As seen in the previous paragraphs, the probability non-trivially depends on various parameters, especially the initial amplitude. Regardless, we would like to examine whether the probability averages to $1/\mathcal{N}$ if the QCD timing and the initial amplitude are both marginalized, even though $\theta_i$ is merely an initial condition rather than an uncertainty. Namely, we aim to provide a precise, zeroth-order statistical understanding of the $Z_\mathcal N$ model. We perform the analysis for $3 \le \mathcal{N} \le 29$ and 12 benchmark masses between $6 \times 10^{-12} \eV \le \maqcd \le 10^{-5} \eV$ equally spaced on logarithmic scale for each $\mathcal{N}$. For each benchmark point, the initial misalignment angle, $\theta_i$, is varied over $|\pi - \theta_i| \in (0,\pi)$ with $256$ points. Furthermore, for every $\theta_i$, the timing at $T_{\rm QCD}$ is varied over an oscillation cycle or up to 2\% of $T_{\rm QCD}$, whichever is smaller, with $256$ points. Fig.~\ref{fig:final_min} illustrates two examples of this analysis, where the upper (lower) panel involves variation over an oscillation cycle (over the full 2\% uncertainty in $T_{\rm QCD}$). We first calculate the probability of solving the strong CP problem among the 256 points for each $\theta_i$. We further calculate the average of these probabilities for points satisfying $|\pi - \theta_i| \in (\theta_c,\pi)$. The result is shown in Fig.~\ref{fig:prob1} with the vertical axis showing $1/P$, where each benchmark point is shown as a square for the specified $\mathcal{N}$ and $\maqcd$. The black line tracks $P = 1/\mathcal{N}$ for reference. If the initial amplitude is not restricted but allowed to vary over all possible values $(0,\pi)$, the resultant average probability is significantly lower and shown in Fig.~\ref{fig:prob_overall}.

As can be seen from Fig.~\ref{fig:prob1}, intermediate $\maqcd$ tends to have $P = 1/\mathcal{N}$ when averaged over initial amplitudes restricted to above $\theta_c$. When the mass is too large, the residual oscillation amplitude/velocity at $T_{\rm QCD}$ too small for the field to reach the CP-conserving minimum. Then the probability becomes zero for all $\theta_i$, explaining the absence of heavy mass points. When the mass is small, the oscillation time scale is not much shorter than $\Delta t_{\rm QCD}$, so a QCD timing variation is not broadly sampling over many phases of the oscillation cycles, resulting in little variation in the final minimum reached. Thus, once $|\pi - \theta_i| > \theta_c$, the probability is nearly unity. It is the intermediate mass regime that allows for sufficient energy to reach the CP-conserving minimum and sufficient cycle sampling to generate a random selection. 

From Fig.~\ref{fig:prob_overall}, we see that for small $\maqcd$, the probability approaches $P = 1/\mathcal{N}$ when averaged over all initial amplitudes. The reason is as follows. For small $\maqcd$, $P \simeq 1$ under the condition that $|\pi - \theta_i| > \theta_c$, but approximately $1/\mathcal{N}$ fraction of the initial angle range satisfies the condition. For intermediate $\maqcd$, fair sampling over multiple oscillation cycles causes the probability to drop below unity even when $|\pi - \theta_i| > \theta_c$. This explains why the probability starts to deviate from $1/\mathcal{N}$ as $\maqcd$ increases. For large $\maqcd$, the ability to reach the CP-conserving minimum decreases due to the small residual velocity at $T_{\rm QCD}$, and we need a more suppressed potential (namely, a larger $\mathcal{N}$) to compensate for it. This explains why the probability increases when $\mathcal{N}$ increases for a given $\maqcd$.

Hence, there is indeed some statistical feature of $P = 1/\mathcal{N}$ in some cases, but again marginalizing over initial angles is not generally physically meaningful. In most of the cases, the probability is nowhere close to $1/\mathcal{N}$. Therefore, we conclude that the probability is not generically equal to $1/\mathcal{N}$ but sensitively depends on the initial amplitude.

\subsection{Dark matter}
\label{subsec:DM}

We start this section with an analytic estimate of the dark matter abundance in the $Z_\mathcal{N}$ model. For $\maqcd \gtrsim 3H_{\rm QCD} \simeq 4 \times 10^{-11} \eV$, oscillations around the minimum at $\theta = \pi$ begin before $T = T_{\rm QCD}$. The oscillation amplitude redshifts between the onset of oscillation and the QCD phase transition according to $T^{3/2}$, and thus the amplitude at $T_{\rm  QCD}$ is given by $\Phi(T_{\rm  QCD})/f_a \simeq  |\pi-\theta_i| (T_{\rm QCD} / T_{\rm osc})^{3/2}$, where $T_{\rm osc}$ is determined by $\maqcd = 3 H(T_{\rm osc})$. If the residual angular velocity $\dot\theta(T_{\rm  QCD}) = \dot\Phi/f_a \simeq \maqcd \times |\pi-\theta_i| (T_{\rm QCD} / T_{\rm osc})^{3/2}$ is larger than $2 m_a/\mathcal{N}$, then kinetic misalignment occurs because the kinetic energy, $\dot\theta^2 f_a^2/2$, is larger than the potential barrier, $2 m_a^2 f_a^2/\mathcal{N}^2$ using Eq.~\eqref{eq:simp_ZNpot}. In this case, the dark matter abundance is analytically given by
\begin{equation}
\label{eq:KMM}
    \frac{\rho}{s} \simeq \left. \frac{\rho}{s} \right|_{T_{\rm trap}} \simeq m_a \left. \frac{ \dot\theta f_a^2}{\mathcal{N} s} \right|_{T_{\rm  QCD}} ,
\end{equation}
where $\rho(T_{\rm trap}) = 2 m_a^2 f_a^2/\mathcal{N}^2$ is the height of the potential barrier at the time when the axion is trapped, $\dot\theta(T_{\rm trap}) \simeq 2 m_a/\mathcal{N}$. In the second equality, we have used the fact that $\dot\theta f_a^2 /s$ is redshift invariant since this corresponds to an approximately conserved $U(1)$ charge~\cite{Co:2019jts} during the rolling phase. To explain the observed dark matter abundance, $\rho_{\rm DM}/s = 0.44 \eV$, the axion mass is constrained to 
\begin{equation}
    f_a \simeq 10^{14} \GeV~ \theta_i^{-4/7}
    \left( \frac{10^{-10} \eV}{m_a} \right)^{ \scalebox{0.9}{$\frac{4}{7}$} } 
    \left( \frac{\mathcal{N}}{25} \right)^{ \scalebox{0.9}{$\frac{4}{7}$} } .
\label{eq:fa_analytic}
\end{equation}
The estimate from this approach is broadly consistent with the latest lattice result in Ref.~\cite{Fasiello:2025ptb} regarding kinetic misalignment for the standard QCD axion.

On the other hand, if $\dot\theta(T_{\rm  QCD}) < 2 m_a/\mathcal{N}$, then the axion is immediately trapped to the nearest minimum, and the abundance is given by
\begin{equation}
    \frac{\rho}{s}  \simeq \left. \frac{\rho}{s} \right|_{T_{\rm QCD}} \simeq \left. \frac{ 2 m_a^2 f_a^2}{\mathcal{N}^2 s} \right|_{T_{\rm QCD}}   .
    \label{eq:MM}
\end{equation}
Note this regime immediately implies that the strong CP problem necessarily cannot be solved for any $\mathcal{N} > 1$ because the field becomes trapped in the minimum closest to $\theta=\pi$.

Lastly, if $\maqcd \lesssim 3H_{\rm QCD} \simeq 4 \times 10^{-11} \eV$, the axion field does not oscillate toward $\theta=\pi$ before $T_{\rm QCD}$, and the conventional misalignment mechanism simply applies.

\begin{figure}
    \centering
    \includegraphics[width=1.0\linewidth]{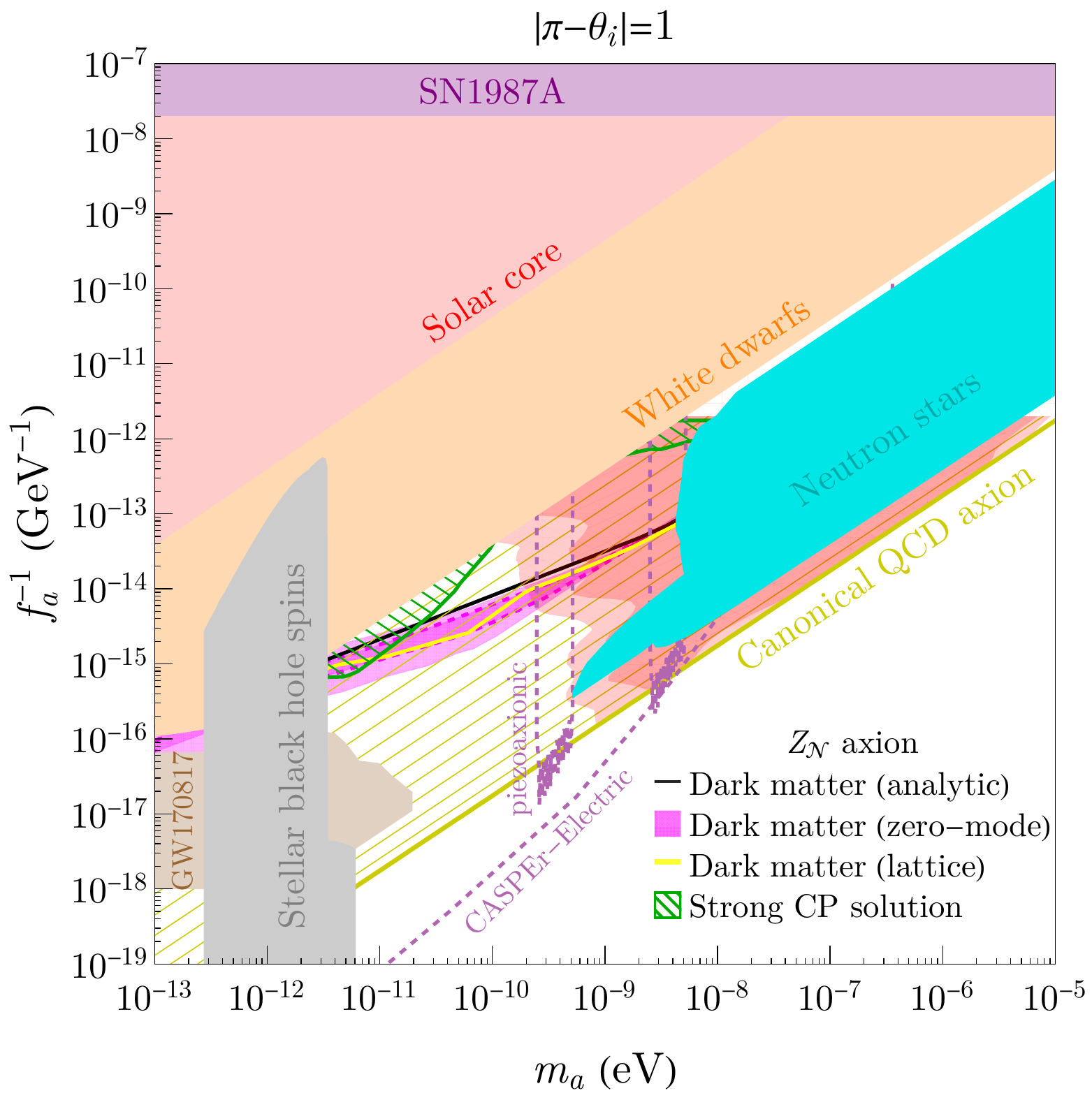}
    \includegraphics[width=1.0\linewidth]{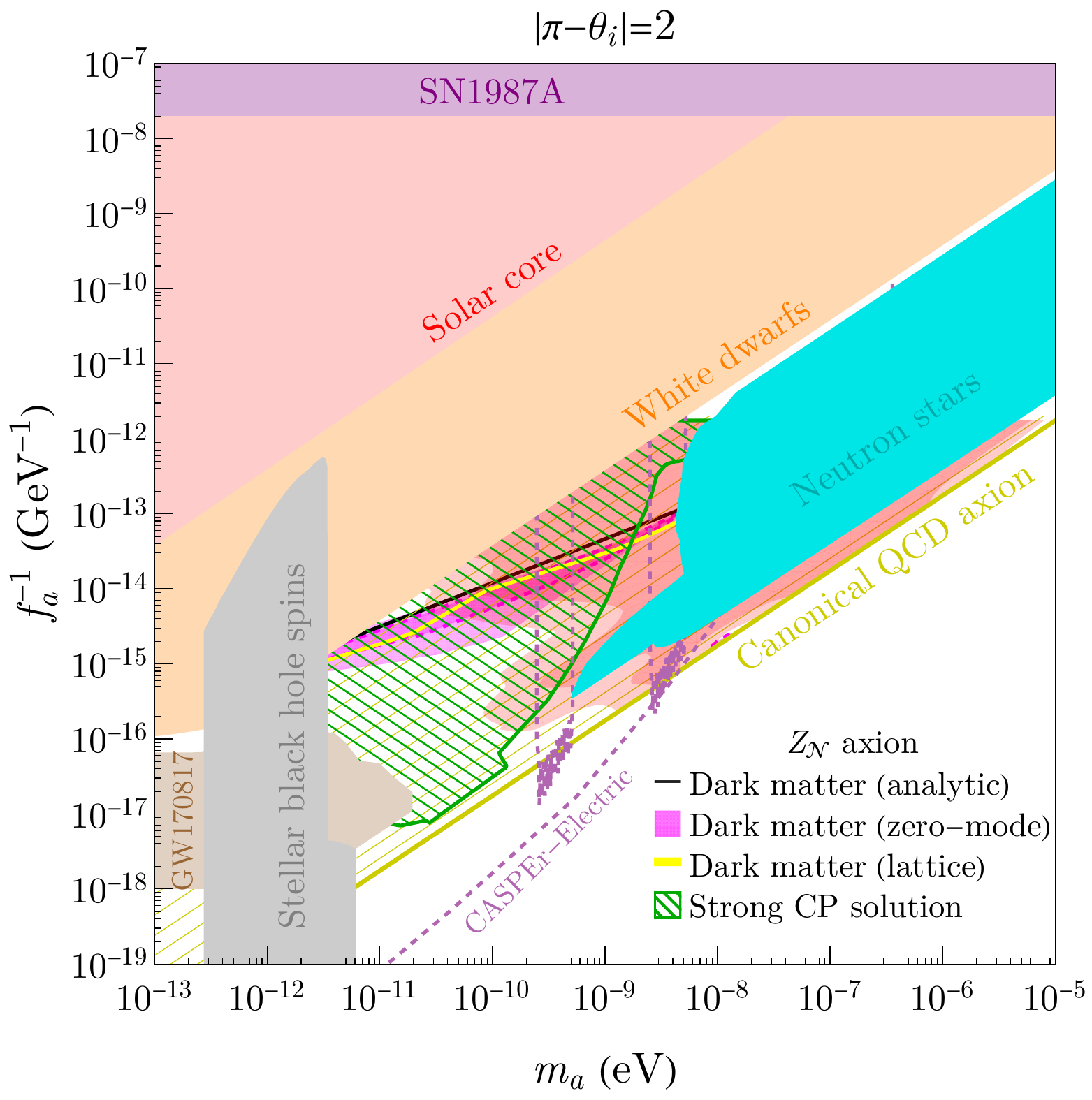}
    \caption{Viable parameter space for the $Z_\mathcal{N}$ QCD axion. The upper and lower panels assume initial amplitudes $|\pi - \theta_i| = 1$ and $2$, respectively. The black line shows the contour of the observed dark matter abundance using the analytic estimates in Eqs.~\eqref{eq:KMM} and~\eqref{eq:MM}. Regions above/below lead to under/overproduction of dark matter. For the numerical analysis involving only the zero mode, dark matter can be accounted for in the magenta-shaded region, where the width reflects the uncertainty of the QCD phase transition temperature. The magenta dashed lines represent the one standard-deviation range. The yellow line shows the dark matter contour from the lattice results. The strong CP problem can be solved in the green hatched region for at least one choice of $T_{\rm QCD}$ within the theoretical uncertainty. The dark/light red-shaded region shows the strength of backreaction according to $\left. \rho_G/\rho_{\rm total} \right |_{\rm max} = 10^{-3}$ and $10^{-9}$, respectively.}
    \label{fig:summary}
\end{figure}

The analytic estimates result in the black lines in Fig.~\ref{fig:summary}, indicating the observed dark matter abundance. The initial amplitude is smaller in the upper panel than the lower, and hence it is more likely to be in the conventional misalignment regime. Indeed, the change of slope of the black line in the upper panel in the large mass regime is due to the transition between conventional and kinetic misalignment. Larger (smaller) $\maqcd$ leads to more (less) redshift suppression in $\dot\theta (T_{\rm QCD})$ and thus a smaller (larger) kinetic energy at $T_{\rm QCD}$, leading up to conventional (kinetic) misalignment. In the lower panel, the entire black line is in the kinetic misalignment regime given in Eq.~\eqref{eq:fa_analytic}. The green-hatched regions are identical to the $\theta_c=1$ and $\theta_c=2$ contours in Fig.~\ref{fig:theta_c}. Inside the green-hatched regions, there is a chance of solving the strong CP problem for a right value of $T_{\rm QCD}$ within the allowed window of the 2\% theoretical uncertainty.

\begin{figure}
    \centering
    \includegraphics[width=1.0\linewidth]{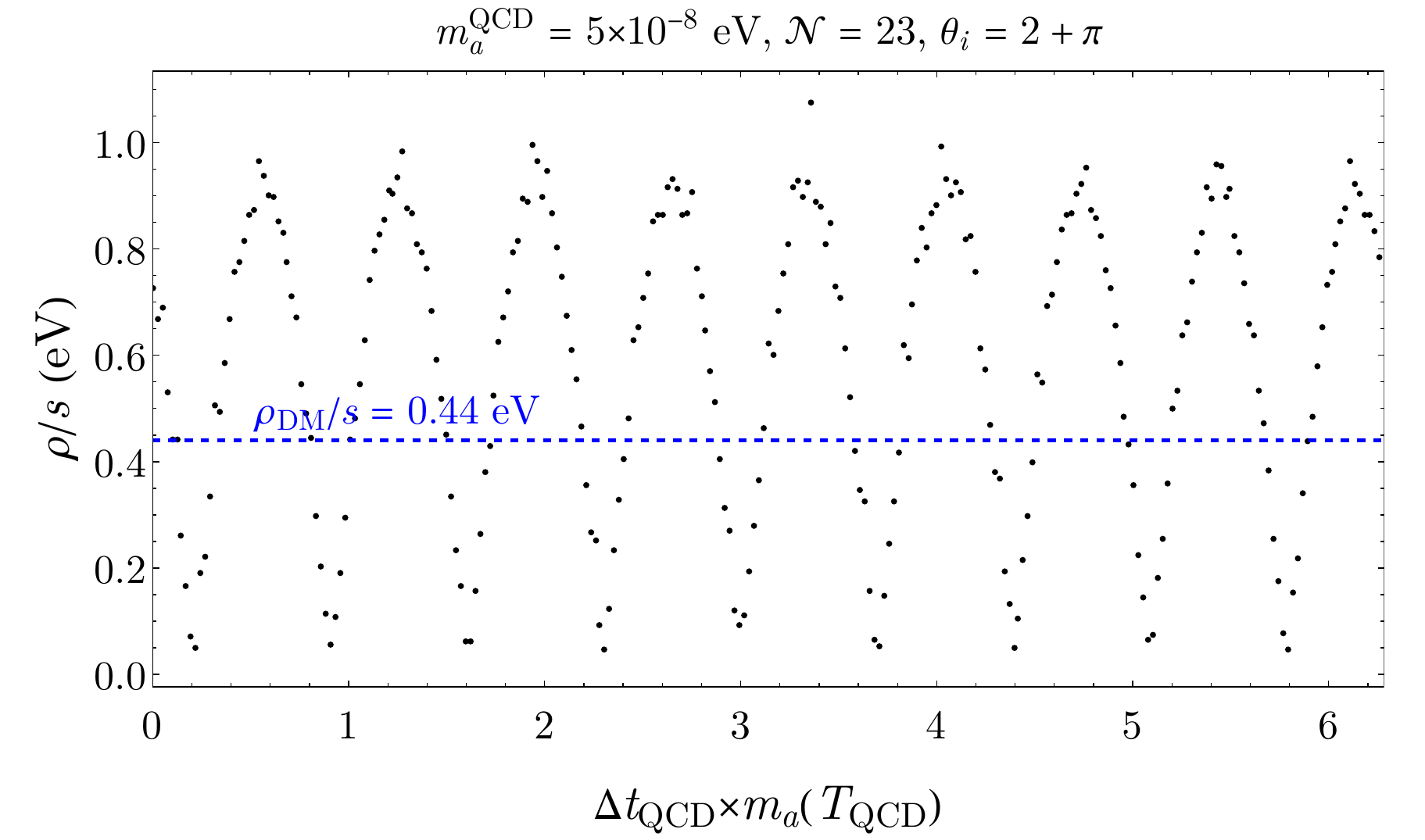}
    \includegraphics[width=1.0\linewidth]{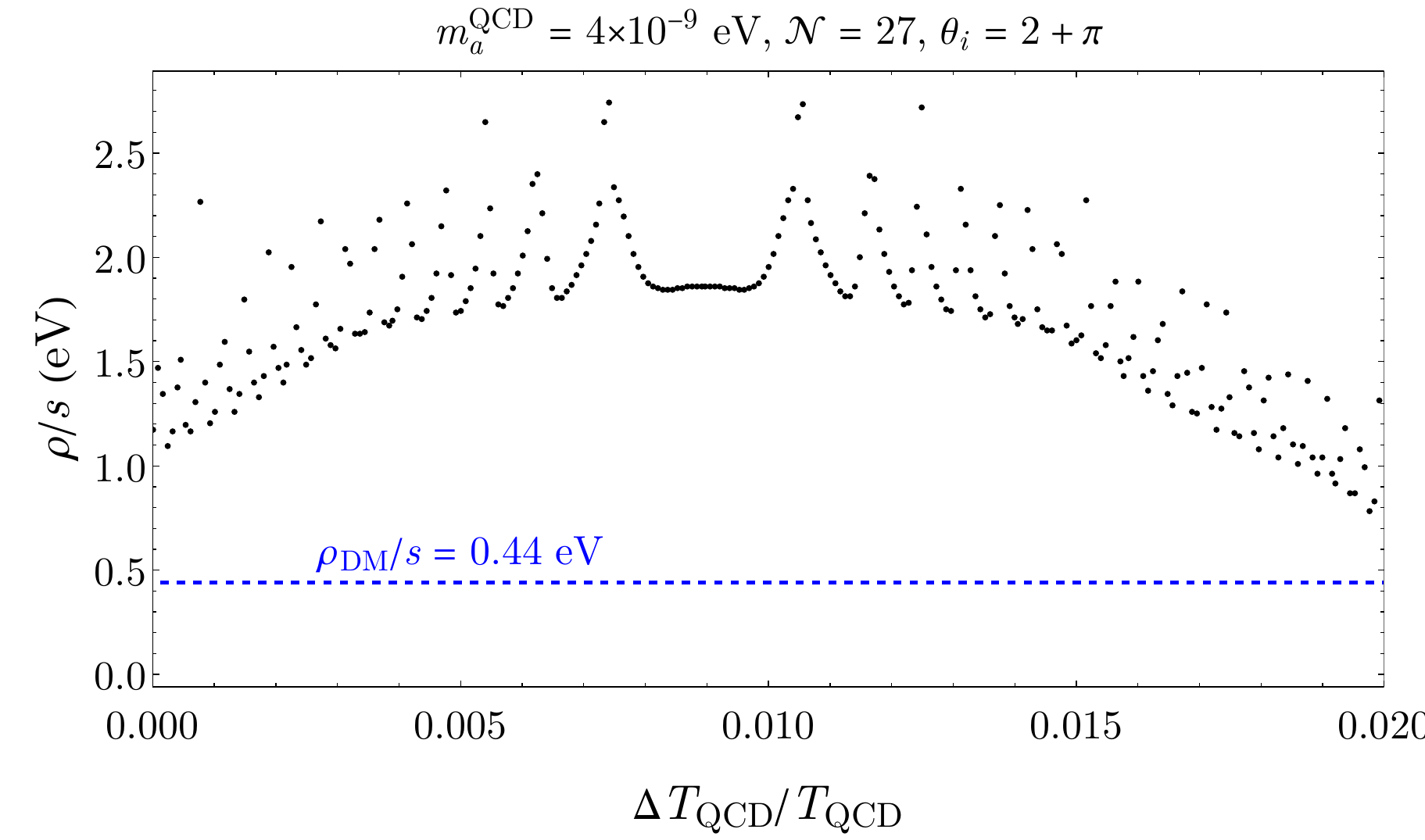}
    \caption{The black dots show the axion abundance as a function of QCD time/temperature variation (upper/lower panel) from the zero-mode analysis with parameters specified at the top axes. The blue dashed line shows the observed dark matter abundance.}
    \label{fig:rho}
\end{figure}

Next we show the results from the numerical zero-mode analysis by the magenta regions in Fig.~\ref{fig:summary}. The region is obtained by varying $T_{\rm QCD}$ in 256 steps over the theoretical uncertainty of 2\% as described previously in Sec.~\ref{subsec:strongCP}. The upper (lower) boundary is set by the maximum (minimum) of the dark matter abundance among the 256 data points. The upper (lower) magenta dashed line is determined by the mean abundance plus (minus) one standard deviation of the 256 data points. The standard deviation becomes small in the large mass limit, indicating the insensitivity to the exact timing of the QCD phase transition. Fig.~\ref{fig:rho} illustrates two examples of this analysis, where the variation is over one oscillation cycle in the upper panel for a larger mass and is over a full 2\% in $T_{\rm QCD}$ in the lower panel for a smaller mass.

Furthermore, we examine the effect of the  temperature dependence of the topological susceptibility. Along the region with the correct observed dark matter abundance, we find a fluctuation of the abundance at the level of $-20\%$ to $+60\%$ if the lattice QCD results~\cite{Borsanyi:2016ksw} are utilized as opposed to the broken power law in Eq.\eqref{eq:Pot_Powerlaw_Tdep}.

On the lattice, we calculate the relic abundance in terms of the dark matter yield,
\begin{equation}
    \frac{\rho}{s}=\frac{\rho_V+\rho_K+\rho_G}{\frac{2 \pi^2}{45}g_{*S}T^3}\,,
    \label{eq:LatticeDMYield}
\end{equation}
where $\rho_V$, $\rho_K$, $\rho_G$ are the potential, kinetic, and gradient components of the energy density each and $g_{*S}$ is the effective number of degrees of freedom in entropy. This method requires running the simulations sufficiently long for all produced particles to become nonrelativistic, ensuring that $\rho/s$ becomes redshift invariant. The final result is shown by the yellow lines, which are consistent with the analytic and zero-mode analyses across a wide range of the parameter space. For the lower panel with $|\pi - \theta_i| = 2$, the lattice results start to yield smaller axion abundances than those of the analytic and zero-mode analyses in the large mass regime. This occurs when PR is very efficient and the backreaction becomes sufficiently strong to completely destroy the zero mode. The dark red-shaded region shows where efficient PR and backreaction are observed. Quantitatively, this region is obtained by requiring the ratio of the gradient energy $\rho_G$ to the total to exceed $10^{-3}$ at some point during the evolution, namely $\left. \rho_G/\rho_{\rm total} \right |_{\rm max} > 10^{-3}$.%
\footnote{We truncate the red regions at the smallest $f_a$ value to which we simulate.} 
Besides, the light red-shaded region, $\left. \rho_G/\rho_{\rm total} \right |_{\rm max} > 10^{-9}$, shows the existence of PR but negligible backreaction. We observe no data points with $\left. \rho_G/\rho_{\rm total} \right |_{\rm max}$ between $10^{-6}$ and $10^{-3}$ in our scan.

\begin{figure}
    \centering
    \includegraphics[width=1.0\linewidth]{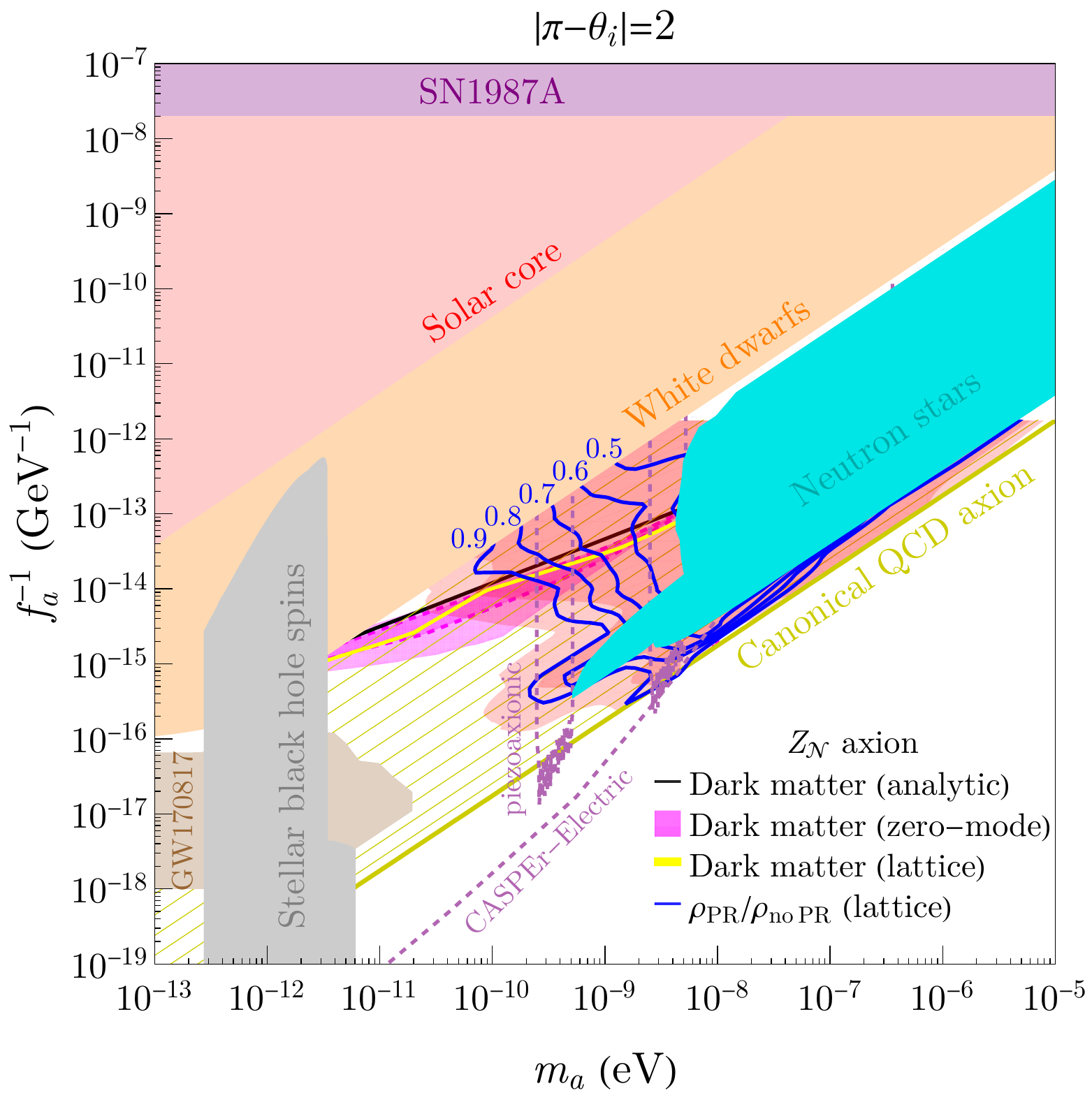}
    \caption{Contours of the non-perturbative correction factor, which we define to be the lattice dark matter relic density, $\rho_{\rm PR}$, which includes non-perturbative PR effects, divided by that from a pure zero-mode evolution, $\rho_{\rm no\;PR}$. }
    \label{fig:nonpertub_correction}
\end{figure}

We obtain the contours of the non-perturbative correction to $\rho/s$ for $|\pi-\theta_i|=2$, shown in Fig.~\ref{fig:nonpertub_correction}, by repeating the 140-point scan with each simulation's spectral window shifted sufficiently below the dynamical range, so as to render any non-perturbative effects invisible on the lattice. Although the lattice is unnecessary to simulate pure zero-mode evolution, such a procedure allows a direct, software-independent comparison of the non-perturbative and zero-mode $\rho/s$ at each benchmark. As expected, contours of the non-perturbative correction roughly trace the boundary between the red-shaded PR regions. As the axion goes deeper into the strong PR domain of the parameter space, this non-perturbative correction becomes correspondingly more significant. The most extreme non-perturbative correction near the lattice dark matter band observed in the allowed parameter space is roughly $0.5$, corresponding to a decrease in the dark matter abundance by a factor of two.

As with the zero-mode $T_{\rm QCD}$-varying analysis done in Sec.~\ref{subsec:strongCP}, secondary lattice simulations for selected benchmark points along the lattice dark matter band in Fig.~\ref{fig:summary} were done in order to determine how non-perturbative dynamics affect the results of the analysis demonstrated in Fig.~\ref{fig:rho}. From this study, we find variations in $\rho/s$ with $T_{\rm QCD}$ in the non-perturbative regime which are in good agreement with the results of the zero-mode analysis, for the selected benchmarks---for more details, as well as the numerical results of this secondary analysis, see App.~\ref{sec:appendix_B}.

The overlap between the green hatched regions and the dark matter lines/regions in Fig.~\ref{fig:summary} serves as our prediction of the parameters for the $Z_{\mathcal{N}}$ QCD axion model.

\section{Summary and Discussion}
\label{sec:summary}
Recently, the $Z_{\mathcal N}$ model \cite{Hook:2018jle} has gained significant attention due to it being the only known framework currently that naturally permits a lighter QCD axion. This model features $\mathcal{N}$ degenerate minima, only one of which ($\theta =0$) solves the strong CP problem---so the probability was hypothesized to be $1/\mathcal{N}$. $Z_{\mathcal N}$ model has also been studied as a candidate for dark matter using perturbative numerical methods \cite{DiLuzio:2021gos}. In this work, we have studied the robustness of the predictions through a combined zero-mode and non-perturbative analysis. Specifically, we attempted to address two questions in the presence of non-perturbative effects: whether the model still solves the strong CP problem and whether it can account for the total abundance of dark matter.

The main results for this work are summarized in Fig.~\ref{fig:summary}. We found a good agreement in the dark matter abundance between the zero mode and lattice simulation when non-perturbative effects are not significant. However, as the non-perturbative effects becomes stronger, we have found order one suppression (up to a factor of two) in the dark matter abundance from the lattice compared to zero-mode analysis. For a small initial amplitude, e.g.~$|\pi-\theta_i|=1$, solutions that satisfy both the strong CP constraint and the observed dark matter abundance appear only in the low-mass region, $m_a \simeq 10^{-11} \eV$ and $f_a \simeq 10^{15} \GeV$. For large amplitude ($|\pi-\theta_i|=2$), the strong CP problem can be solved in a wide range of parameter space. However, requiring the model to simultaneously solve the strong CP problem and account for the observed dark matter abundance yields a relatively sharp correlation between $m_a$ and $f_a$ parameter space. The relation derived from an analytic estimate is given in Eq.~\eqref{eq:fa_analytic}, and as can be seen in Fig.~\ref{fig:summary}, full numerical analyses generically agree but $f_a$ can be larger by a factor of a few due to dynamical effects. These predictions are within the reach of future axion experiments such as CASPEr-Electric. In addition, we find that the probability of relaxing into the CP-conserving minimum can differ appreciably from $1/\mathcal N$ once realistic dynamical effects are taken into account.

Despite these advances in the comprehensive understanding of the model, several key questions remain unanswered. Primarily, a more detailed domain wall analysis is worth conducting. Specifically, we have found that the finite size domain walls do form even in our case of pre-inflationary PQ breaking due to large fluctuations produced. Since parametric resonance occurs only after trapping in the allowed parameter space, the total axion energy density is insufficient to cause a large field fluctuation. Therefore, we do not expect stable, infinite domain walls to form. We have confirmed that only two neighboring minima next to the final minimum in which we settle are occupied. These domain walls become much smaller than the horizon size at the end of the simulation. Secondly, it would be interesting to investigate the formation of oscillons and track their dynamics in detail. These structures could exhibit rich observational signatures.

\section*{Acknowledgment}
R.C.~would like to thank David Cyncynates for useful discussions. O.L.~is grateful to Javier Redondo and Alexandros Papageorgiou for insightful discussions, and to the organizers and sponsors of the 2025 Axions in Stockholm conference, for providing an environment conducive to these interactions. The authors would like thank Anson Hook for discussions. This work was supported by the Department of Energy under Grant No.~DE-SC0025611 at Indiana University. For facilitating portions of this research, R.C. wishes to acknowledge the Center for Theoretical Underground Physics and Related Areas (CETUP*), the Institute for Underground Science at Sanford Underground Research Facility (SURF), and the South Dakota Science and Technology Authority for hospitality and financial support, as well as for providing a stimulating environment when this work was finalized. This research was supported in part by Lilly Endowment, Inc., through its support for the Indiana University Pervasive Technology Institute.

\appendix
\section{Lattice Analysis Parameters}
\label{sec:appendix_A}
In order to save time on the lattice, we begin the field evolution by solving the decoupled zero mode numerically up to what becomes the initial time on the lattice. The stitching of the field evolution between the decoupled zero mode and the lattice is done at a time 100 times earlier than the QCD phase transition, $t_{\rm QCD}/100$. This approach is justified by the fact that the field evolution begins in the harmonic regime of the potential, $\vert \pi - \theta_i \vert = \{1,\,2\}$, where the evolution takes place entirely in the zero mode. From here we can determine our parameters for each benchmark point at $t_{\rm QCD}/100$. In \CLns, this includes $\tilde H(t_{\rm QCD}/100)$, $\phi(t_{\rm QCD}/100)$, and $\dot\phi(t_{\rm QCD}/100)$. During radiation domination, the Hubble rate is $\tilde H=1/(2m_a^{\rm QCD}t)$, where $\tilde H= H/m_a^{\rm QCD}$ is the dimensionless Hubble parameter and $t$ is physical time. Hence $t_{\rm QCD}=1/(2m_a^{\rm QCD}\tilde H_{\rm QCD})$, and $t$ and $T$ are related by $t(T)=\sqrt{45/(2 \pi ^2 g_*(T))} M_{\rm Pl}/T^2$ with $M_{\rm Pl}$ the reduced Planck scale. Although the form of $V_{\mathcal{N}}(\phi)$ produces computationally significant differences in the parameters evaluated at $t_{\rm QCD}/100$ for different $\mathcal{N}$, we provide an example of the above $t_{\rm QCD}/100$ parameters for $\mathcal{N}=13$ in Table~\ref{tab:init_cond_table}, to give the reader some intuition for how these parameters vary for each $m_a^{\rm QCD}$.
\begin{table}[ht]
\centering
\begin{tabular}{|c|c|c|c|}
\hline
\multicolumn{4}{|c|}{$\mathcal{N}=13,\ \vert \pi - \theta_i \vert = 1$} \\
\hline
$m_a^{\rm QCD}$ (eV) & $\tilde H({t_{\rm QCD}/100})$ & $\theta(t_{\rm QCD}/100)$ & $\dot\theta(t_{\rm QCD}/100) / m_a^{\rm QCD}$ \\
\hline
$8.00\times 10^{-11}$ & $1.77\times 10$  & 4.140 & $-1.65\times 10^{-2}$ \\
$2.95\times 10^{-10}$ & $4.79$  & 4.140 & $-6.16\times 10^{-2}$ \\
$1.09\times 10^{-9}$  & $1.30$  & 4.100 & $-2.26\times 10^{-2}$ \\
$4.00\times 10^{-9}$  & $3.53\times 10^{-1}$  & 3.570 & $-7.38\times 10^{-1}$ \\
$1.47\times 10^{-8}$  & $9.58\times 10^{-2}$  & 3.310 & $-2.05\times 10^{-1}$ \\
$5.43\times 10^{-8}$  & $2.60\times 10^{-2}$  & 3.180 & $\,1.08\times 10^{-1}$ \\
$2.00\times 10^{-7}$  & $7.06\times 10^{-3}$  & 3.130 & $\,3.96\times 10^{-2}$ \\
$7.37\times 10^{-7}$  & $1.92\times 10^{-3}$  & 3.130 & $\,1.30\times 10^{-2}$ \\
$2.71\times 10^{-6}$  & $5.20\times 10^{-4}$  & 3.140 & $-2.73\times 10^{-3}$ \\
$1.00\times 10^{-5}$  & $1.41\times 10^{-4}$  & 3.140 & $-2.24\times 10^{-3}$ \\
\hline
\end{tabular}
\begin{tabular}{|c|c|c|c|}
\hline
\multicolumn{4}{|c|}{$\mathcal{N}=13,\ \vert \pi - \theta_i \vert = 2$} \\
\hline
$m_a^{\rm QCD}$ (eV) & $\tilde H({t_{\rm QCD}/100})$ & $\theta(t_{\rm QCD}/100)$ & $\dot\theta(t_{\rm QCD}/100)/ m_a^{\rm QCD}$ \\
\hline
$8.00\times 10^{-11}$ & $1.77\times 10$  & 5.140 & $-1.18\times 10^{-2}$ \\
$2.95\times 10^{-10}$ & $4.79$  & 5.140 & $-4.40\times 10^{-2}$ \\
$1.09\times 10^{-9}$  & $1.30$  & 5.110 & $-1.64\times 10^{-2}$ \\
$4.00\times 10^{-9}$  & $3.53\times 10^{-1}$  & 4.680 & $-6.94\times 10^{-1}$ \\
$1.47\times 10^{-8}$  & $9.58\times 10^{-2}$  & 3.360 & $\,7.02\times 10^{-1}$ \\
$5.43\times 10^{-8}$  & $2.60\times 10^{-2}$  & 2.960 & $-7.80\times 10^{-2}$ \\
$2.00\times 10^{-7}$  & $7.06\times 10^{-3}$  & 3.150 & $-1.18\times 10^{-1}$ \\
$7.37\times 10^{-7}$  & $1.92\times 10^{-3}$  & 3.150 & $-3.95\times 10^{-2}$ \\
$2.71\times 10^{-6}$  & $5.20\times 10^{-4}$  & 3.150 & $\,4.25\times 10^{-3}$ \\
$1.00\times 10^{-5}$  & $1.41\times 10^{-4}$  & 3.140 & $\,5.58\times 10^{-3}$ \\
\hline
\end{tabular}
\caption{Initial lattice conditions for $\mathcal{N}=13$ and $\vert \pi - \theta_i \vert = \{1,2\}$ for each benchmark point. Deviations in $\theta(t_{\rm QCD}/100)$ and $\dot{\theta}(t_{\rm QCD}/100)$ from these values can occur up to order $\mathcal{O}(10^{-1})$ for different $\mathcal N$. $\tilde H_{t_{\rm QCD}/100} \equiv H(t_{\rm QCD}/100)/\maqcd$ is independent of $\mathcal{N}$ for fixed $m_a^{\rm QCD}$.}
\label{tab:init_cond_table}
\end{table}

We consider a region of the parameter space still unconstrained by experiment, simulating axions with $8\times10^{-11} < m_a^{\rm QCD} < 1 \times10^{-6}$ eV, for $\mathcal{N}=3,5,...,29$. The bounds defining this region are astrophysical constraints from neutron star cooling~\cite{Gomez-Banon:2024oux,Kumamoto:2024wjd}, white dwarf composition~\cite{Balkin:2022qer}, stellar black hole spins~\cite{Baryakhtar:2020gao}, and axion-nucleon couplings from GW170817~\cite{Zhang:2021mks}. We conduct a scan of 140 benchmark points within this region composed of 10 logarithmically-spaced $m_a^{\rm QCD}$. Each benchmark is tested for two initial misalignment angles, $\vert \pi - \theta_i \vert = \{1,\,2\}$.

There are a number of computational parameters that need to be specified for the lattice simulation, including the number of lattice sites, $N$, momentum infrared (IR) cutoff, $k_{\rm IR}$, and the simulation time step, $d\tilde\eta$, where $\eta$ is conformal time, and $\tilde\eta=m_a^{\rm QCD}\eta$ is dimensionless conformal time. In the allowed region of the parameter space considered in this work, we use $N=256^3$ for each benchmark point. For each $m_a^{\rm QCD}$, $d\tilde\eta$ varies. In the previous paragraph we showed that aside from $T_{\rm QCD}$, which we hold fixed in the initial benchmark scan, $m_a^{\rm QCD}$ uniquely determines $\tilde H(t_{\rm QCD}/100)$. In particular, a larger $m_a^{\rm QCD}$ leads to a correspondingly smaller $\tilde H(t_{\rm QCD}/100)$. In \CLns, the field evolves in $\tilde\eta$. During radiation domination, $\tilde\eta$ governs the evolution of $a$ according to the relation, $a=\tilde H(t_{\rm QCD}/100)\tilde\eta+1$. Thus, larger $m_a^{\rm QCD}$ results in a field which evolves more slowly in $\tilde\eta$, and the simulations can be run with a larger timestep. For this work, $0.0005<d\tilde\eta <0.05$. In addition to specifying the IR cutoff, $k_{\rm IR}$ determines the width of the spectral bins and, together with $N$, sets the spectral range of a simulation by determining the largest $k$-mode detectable by the lattice, $k_{\rm max}=\sqrt{3}Nk_{\rm IR}/2$. The best value of $k_{\rm IR}$ for each benchmark point must not only define a spectral window that captures the relevant dynamics, but also provide sufficient ultraviolet (UV) resolution, so as to avoid contamination from unphysical Nyquist-scale effects. We begin by choosing a value such that the principal resonance peak, $k/a \approx 1.13m_a$, is centered in the spectral range. Spectral data on the lattice is given in physical $k$. Due to the fact that scale factor when PR is strongest varies for each benchmark point, $k_{\rm IR}$ must be adjusted to ensure that the simulation simultaneously captures the full set of dynamically important modes, while maintaining the necessary UV resolution. 

Convergence tests of the above parameters were performed to minimize unphysical lattice effects, including Nyquist effects, on the quantities presented in this work. In particular, our lattice DM abundances are converged to percent-level at $N=256^3$ relative to $N=512^3$.

In addition to specifying the above computational parameters, we take the initial conditions for the fluctuations of the field $\delta \phi_k / f_a$ to be $\mathcal{O}(10^{-5})$, in agreement with the adiabatic perturbation observed in CMB measurements~\cite{Planck:2018vyg}.

\section{Secondary Lattice Simulations}
\label{sec:appendix_B}

Secondary simulations for $\vert \pi - \theta_i \vert = 2$ were done to study the effect of the uncertainty in $T_{\rm QCD}$ on dark matter and the strong CP problem with $T_{\rm QCD}$ raised in 13 equal increments. When the temperature increase during one field oscillation cycle is smaller than the maximum $2\%$ uncertainty, the increments are done over this cycle. If not, the increments are spaced over the full $2\%$ uncertainty, as discussed in Sec.~\ref{subsec:strongCP}. These simulations were done for four benchmark points, characterized by $(m_a^{\rm QCD}\,[\mathrm{eV]},\mathcal{N})$: $(2.95\times10^{-10},9)^{(1)}$, $(1.47\times10^{-8},15)^{(2)}$, $(7.37 \times10^{-7},21)^{(3)}$, and $(7.37 \times10^{-7},27)^{(4)}$. Points $(1)$ and $(2)$ were selected to confirm that the $\theta_c$ determination, considered in Sec.~\ref{subsec:strongCP} with the zero-mode analysis, remains valid in the non-perturbative analysis for the parameter space region considered in this work. Point $(3)$ was studied further as it is a benchmark point in the non-perturbative regime for which the observed dark matter abundance is nearly satisfied, and it did not solve the Strong CP problem in the initial benchmark scan. Point $(4)$, also in the non-perturbative regime, was studied further for the inverse reason---this benchmark point was able to solve the strong CP problem, and was not close to reproducing the observed dark matter abundance. In the case of both $(3)$ and $(4)$, allowing for the uncertainty in $T_{\rm QCD}$ did not allow for the solution of both problems. In the initial lattice scan, points $\{(1),(2),(3),(4)\}$ generated $\rho/s=\{1.878\times10^4,24.91,1.423 \times 10^{-1},1.167\times10^{-2}\}\,\mathrm{eV}$, respectively. The secondary lattice simulations for these points demonstrated maximum deviations from these values of $\{0\%,+179\%,+50\%,-90.6\%\}$.

\bibliography{ZN} 
\end{document}